# Dyadic groups, dyadic trees and symmetries in long nucleotide sequences

Sergey V. Petoukhov

Head of Laboratory of Biomechanical System, Mechanical Engineering Research Institute
of the Russian Academy of Sciences, Moscow
spetoukhov@gmail.com, petoukhov@imash.ru,
http://symmetry.hu/isabm/petoukhov.html, http://petoukhov.com/

**Abstract:** The conception of multi-alphabetical genetics is considered. Matrix forms of the representation of the multi-level system of molecular-genetic alphabets have discovered algebraic properties of this system. These properties are connected with well-known notions of dyadic groups and dyadic-shift matrices. Matrix genetics shows relations of the genetic alphabets with some types of hypercomplex numbers including dual numbers and bicomplex numbers together with their extensions. A possibility of new approach is mentioned to simulate genetically inherited phenomena of biological spirals and phyllotaxis laws on the base of screw theory and Fibonacci matrices. Dyadic trees for sub-sets of triplets of the whole human genome are constructed. The notion is put forward about square matrices with internal complementarities on the base of genetic matrices. Initial results of the study of such matrices are described. Our results testify that living matter possesses a profound algebraic essence. They show new promising ways to develop algebraic biology.

## Contents

### 1. Introduction

Compositional features of nucleotide sequences are studied for decades because of their importance for bioinformatics and theoretical biology. Many of them are devoted to the

Chargaff's second parity rule [Bell & Forsdyke, 1999; Chen & Zhao, 2005; Dong, Cuticchia, 2001; Forsdyke & Bell, 2004; Forsdyke, 1995, 2002; Mitchell & Bride, 2006; Perez, 2010; Prahbu, 1993; Qi & Cuticchia, 2001; Yamagishi & Herai, 2011]. Chargaff's first parity rule states that the frequency of A is equal to that of T and the frequency of C is equal to that of G in double-stranded DNA. Watson and Crick were acquainted with this rule, and used it to support their famous DNA double-helix structure model [Watson & Crick, 1953]. Chargaff also perceived that the parity rule approximately holds in the single-stranded DNA segment. This last rule is known as Chargaff's second parity rule, and although it is not well understood, it has been confirmed in several organisms [Mitchell & Bride, 2006]. Originally, the second rule is meant to be valid only to mononucleotide frequencies. But, it occurs that oligonucleotide frequencies follow a generalized Chargaff's second parity rule where the frequency of an oligonucleotide (that is a k-plet or a k-word)) is approximately equal to its complement reverse oligonucleotide frequency [Prahbu, 1993]. This is known in the literature as the Symmetry Principle [Yamagishi & Herai, 2011]. Chargaff's rules are important because they point to a kind of "grammar of biology", a set of hidden rules that govern the structure of DNA.

This article is devoted to further study of symmetries in long nucleotide sequences. This study is based on mathematical formalisms of dyadic groups, modulo-2 addition and matrix genetics from the author's article [Petoukhov, 2012], and this new article can be considered as an independent continuation of the previous one.

The author believes that the way to store and process information in a molecular-genetic system is significantly different from the usual method of storing and processing information in conventional computers and in literature sources. For example, an ordinary book is usually written on the basis of a single alphabet underlying the language (in this example we speak about Indo-European languages each of which possesses its own alphabet). A person reads different parts of the book by means of the same knowledge of a particular alphabet and rules of reading, which he has studied in school. By analogy in computer technology, different fragments of a text written on a CD-ROM are read by means of one program. In short, the following principle is realized: "a piece of text - only one program of its reading", or "a text is a mono-linguistic object".

Any molecular-genetic sequence is a polyatomic structure, which contains carbon, nitrogen, oxygen, hydrogen, purines, pyrimidines and so on. Interpreting of the presence or absence of certain of these different components of the sequence, for example, as the binary digits 0 and 1, one can implement entirely different binary texts on the same molecular sequence. In other words, the genetic text is multi-lingual, since each of its fragments thereof may be represented by different binary texts. This allows implementing in genetic informatics quite another principle of storing and processing information: "a genetic sequence - a multi-lingual text", or "a piece of text - various programs for its reading parallely". This principle or concept can be termed briefly as "multi-lingual genetics" or "multi-alphabetical genetics". In some extend a similar principle is well-known in music where two different alphabets (the alphabet of sound frequencies of musical notes and the alphabet of durations of musical notes) are combined in musical works. In polyphonic music, its "lingual" structure is further complicated because each of the many instruments performs its own musical party (its own "text"), although the entire orchestra provides a single musical product to bring a powerful emotional response from the audience. Something similar implemented by nature in the genetic system. One might think that the fundamentals of music are closer to the fundamentals of the genetic system than individual linguistic languages with their mono-alphabetic base. Author's research is focused on the study of the multilingual nature of the genetic texts by means of matrix representation of the multi-level system of genetic alphabets [Petoukhov, 2008a, 2008b, 2011a,b, 2012; Petoukhov, He, 2010].

## 2. Dyadic groups and matrices of dyadic shifts

Modulo-2 addition is utilized broadly in the theory of discrete signal processing as a fundamental operation for binary variables. By definition, the modulo-2 addition of two numbers written in binary notation is made in a bitwise manner in accordance with the following rules:

$$0 + 0 = 0,\ 0 + 1 = 1,\ 1 + 0 = 1,\ 1 + 1 = 0 \qquad (1)$$

For example, modulo-2 addition of two binary numbers 110 and 101, which are equal to 6 and 5 respectively in decimal notation, gives the result $110 \oplus 101 = 011$, which is equal to 3 in decimal notation ($\oplus$ is the symbol for modulo-2 addition). The set of binary numbers

$$000,\ 001,\ 010,\ 011,\ 100,\ 101,\ 110,\ 111 \qquad (2)$$

forms a diadic group with 8 members, in which modulo-2 addition serves as the group operation [Harmuth, 1989]. By analogy dyadic groups of binary numbers with $2^n$ members can be presented. The distance in this symmetry group is known as the Hamming distance. Since the Hamming distance satisfies the conditions of a metric group, the dyadic group is a metric group. The modulo-2 addition of any two binary numbers from (2) always gives a new number from the same series. The number 000 serves as the unit element of this group: for example, $010 \oplus 000 = 010$. The reverse element for any number in this group is the number itself: for example, $010 \oplus 010 = 000$. Each member from (2) possesses its inverse-symmetrical partner (or a mating number), which arises if the binary symbol of the member is transformed by the inverse replacements 0→1 and 1→0. For example, binary numbers 010 and 101 give an example of such pair of mating numbers.

The series (2) is transformed by modulo-2 addition with the binary number 001 into a new series of the same numbers:

$$001,\ 000,\ 011,\ 010,\ 101,\ 100,\ 111,\ 110 \qquad (3)$$

Such changes in the initial binary sequence, produced by modulo-2 addition of its members with any binary numbers (2), are termed dyadic shifts [Ahmed and Rao, 1975; Harmuth, 1989]. If any system of elements demonstrates its connection with dyadic shifts, it indicates that the structural organization of its system is related to the logic of modulo-2 addition. The article shows additionally that the structural organization of genetic systems is related to logic of modulo-2 addition.

By means of dyadic groups a special family of $(2^n \ast 2^n)$-matrices can be constructed which are termed "matrices of dyadic shifts" and which are used widely in technology of discrete signal processing [Ahmed, Rao, 1975; Harmuth, 1977, §1.2.6]. Figure 1 shows examples of matrices of dyadic shifts (or briefly DS-matrices). In these matrices their rows and columns are numerated by means of binary numbers of an appropriate dyadic group. All matrix cells are numerated by means of binary numbers of the same dyadic group in such way that a binary numeration of each cell is a result of modulo-2 addition of binary numerations of its column and its row. For example, the cell from the column 110 and the row 101 obtains the binary numeration 011 by means of such addition. Such numerations of matrix cells are termed "dyadic-shift numerations" (or simply "dyadic numeration").

|      | 0 | 1 |
|------|---|---|
| 0    | 0 | 1 |
| 1    | 1 | 0 |

;

|        | 00 (0) | 01 (1) | 10 (2) | 11 (3) |
|--------|--------|--------|--------|--------|
| 00 (0) | 00 (0) | 01 (1) | 10 (2) | 11 (3) |
| 01 (1) | 01 (1) | 00 (0) | 11 (3) | 10 (2) |
| 10 (2) | 10 (2) | 11 (3) | 00 (0) | 01 (1) |
| 11 (3) | 11 (3) | 10 (2) | 01 (1) | 00 (0) |

|         | 000 (0) | 001 (1) | 010 (2) | 011 (3) | 100 (4) | 101 (5) | 110 (6) | 111 (7) |
|---------|---------|---------|---------|---------|---------|---------|---------|---------|
| 000 (0) | 000 (0) | 001 (1) | 010 (2) | 011 (3) | 100 (4) | 101 (5) | 110 (6) | 111 (7) |
| 001 (1) | 001 (1) | 000 (0) | 011 (3) | 010 (2) | 101 (5) | 100 (4) | 111 (7) | 110 (6) |
| 010 (2) | 010 (2) | 011 (3) | 000 (0) | 001 (1) | 110 (6) | 111 (7) | 100 (4) | 101 (5) |
| 011 (3) | 011 (3) | 010 (2) | 001 (1) | 000 (0) | 111 (7) | 110 (6) | 101 (5) | 100 (4) |
| 100 (4) | 100 (4) | 101 (5) | 110 (6) | 111 (7) | 000 (0) | 001 (1) | 010 (2) | 011 (3) |
| 101 (5) | 101 (5) | 100 (4) | 111 (7) | 110 (6) | 001 (1) | 000 (0) | 011 (3) | 010 (2) |
| 110 (6) | 110 (6) | 111 (7) | 100 (4) | 101 (5) | 010 (2) | 011 (3) | 000 (0) | 001 (1) |
| 111 (7) | 111 (7) | 110 (6) | 101 (5) | 100 (4) | 011 (3) | 010 (2) | 001 (1) | 000 (0) |

Figure 1. The examples of matrices of dyadic shifts. Parentheses contain expressions of the numbers in decimal notation.

Now let us turn to symmetries in molecular characteristics of the 4-letter alphabet of nitrogenous bases in DNA: adenine A, cytosine C, guanine G and thymine T. These symmetries provide the existence of its binary sub-alphabets. The four letters (or the four nitrogenous bases) of the genetic alphabet represent specific poly-nuclear constructions with the special biochemical properties. The set of these four constructions is not absolutely heterogeneous, but it bears the substantial symmetric system of distinctive-uniting attributes (or, more precisely, pairs of "attribute-antiattribute"). This system of pairs of opposite attributes divides the genetic four-letter alphabet into various three pairs of letters by all three possible ways; letters of each such pair are equivalent to each other in accordance with one of these attributes or with its absence.

Really, the system of such attributes divides the genetic four-letter alphabet into various three pairs of letters, which are equivalent from a viewpoint of one of these attributes or its absence: 1) C = T & A = G (according to the binary-opposite attributes: "pyrimidine" or "non-pyrimidine", that is purine); 2) A = C & G = T (according to the attributes "keto" or "amino" [Karlin, Ost, Blaisdell, 1989]; 3) C = G & A = T (according to the attributes: three or two hydrogen bonds are materialized in these complementary pairs). The possibility of such division of the genetic alphabet into three binary sub-alphabets is known from the work [Karlin, Ost, Blaisdell, 1989]. We will utilize these known sub-alphabets by means of a new method in the field of matrix genetics. We will attach appropriate binary symbols "0" or "1" to each of the genetic letters based on these sub-alphabets. Then we will use these binary symbols for binary numbering the columns and the rows of genetic $(2^n*2^n)$-matrices of Kronecker families.

Let us mark these three kinds of binary-opposite attributes by numbers $N = 1, 2, 3$ and ascribe to each of the four genetic letters the symbol "$0_N$" (the symbol "$1_N$") in case of presence (of absence correspondingly) of the attribute under number "$N$" to this letter. As a result we obtain the following representation of the genetic four-letter alphabet in the system of its three "binary sub-alphabets corresponding to attributes" (Fig. 2).

|   | Symbols of a genetic letter from a viewpoint of a kind of the binary-opposite attributes | C | A | G | T |
|---|---|---|---|---|---|
| №1 | $0_1$ – pyrimidines (one molecular ring) $1_1$ – purines (two molecular rings) | $0_1$ | $1_1$ | $1_1$ | $0_1$ |
| №2 | $0_2$ – amino $1_2$ – keto | $0_2$ | $0_2$ | $1_2$ | $1_2$ |
| №3 | $0_3$ – a letter with 3 hydrogen bonds $1_3$ – a letter with 2 hydrogen bonds | $0_3$ | $1_3$ | $0_3$ | $1_3$ |

Figure 2. Three binary sub-alphabets according to three kinds of binary-opposite attributes in the set of nitrogenous bases C, A, G, T. The scheme on the right side explains graphically the symmetric relations of equivalence between the pairs of letters from the viewpoint of the separate attributes 1, 2, 3 (from [Petoukhov, 2008a, 2012; Petoukhov, He, 2010])

The table on Figure 2 shows that, on the basis of each kind of the attributes, each of the letters A, C, G, T possesses three "faces" or meanings in the three binary sub-alphabets. On the basis of each kind of the attributes, the genetic four-letter alphabet is curtailed into the two-letter alphabet. For example, on the basis of the first kind of binary-opposite attributes we have (instead of the four-letter alphabet) the alphabet from two letters $0_1$ and $1_1$, which one can name "the binary sub-alphabet to the first kind of the binary attributes".

Accordingly, any genetic message as a sequence of the four letters C, A, G, T consists of three parallel and various binary texts or three different sequences of zero and unit (such binary sequences are used at storage and transfer of the information in computers). Each from these parallel binary texts, based on objective biochemical attributes, can provide its own genetic function in organisms. According to our data, the genetic system uses the possibility to read triplets from the viewpoint of different binary sub-alphabets. Taking into account these three sub-alphabets, each n-plet is a carrier of one of the binary numbers of an appropriate dyadic group. For example from the viewpoint of the first sub-alphabet on Figure 2 (C=T=0, A=G=1), the triplet CAG is a carrier of the dyadic number 011. From the viewpoint of the second sub-alphabet (C=A=0, G=T=1) the same triplet CAG is a carrier of the dyadic number 001. From the viewpoint of the third sub-alphabet (C=G=0, A=T=1) the same triplet CAG is a carrier of the dyadic number 010. Below we will analyze genetic (8*8)-matrices of 64 triplets from the viewpoint of these sub-alphabets to reveal numerical symmetries in long sequences of triplets and to reveal an existence of dyadic trees for such sequences by analogy with dichotomous trees in linguistic alphabets.

### 3. The first variant of dyadic numerations of triplets and dyadic trees of triplet frequencies in long nucleotide sequences

Alphabets play a basic role in communication technologies. In any communication system of "transmitter-receiver" the receiver always knows the alphabet of signals, which are used by the transmitter. In Indo-European languages each alphabet has a complex multi-level structure because it contains sets of vowels and consonants where the set of vowels is divided into sub-sets of short and long sounds, and the set of consonants is divided into subsets of voiced and voiceless consonants, etc. (Fig. 2). Quantities of members in all of these parts of linguistic alphabets are not interrelated by means of known regularities of algebraic connections. We have discovered that the situation in the multi-level system of genetic alphabets is quite different: many parts of this system are closely interrelated by means of deep algebraic regularities and formalisms, which are well known in communication technologies [Petoukhov, 2008a, 2011a,b; 2012; Petoukhov, He, 2010].

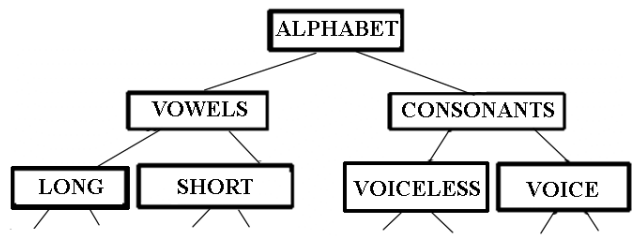

Figure 3. The dichotomous scheme of alphabetical trees in Indo-European languages

This section shows an example of analogical dichotomous schemes in long DNA sequences of 64 triplets. It shows also one of examples of hidden connections between regularities of nucleotide sequences and dyadic groups.

Let us consider the genetic (8*8)-matrix of 64 triplets, which is the third Kronecker power of the (2*2)-matrix [C A; G T] (Figure 4). Here we continue analyses of Kronecker families of genetic matrices from works [Petoukhov, 2008a, 2008b, 2012; Petoukhov, He, 2010]. In accordance with the first sub-alphabet on Figure 2 one can denote each pyrimidine by binary number "0" (C=T=0) and each purine by binary number "1" (A=G=1). In such case each triplet obtains its corresponding binary numeration depending on its three nitrogenous bases. For example the triplet CAG obtains the binary numeration 011. The whole set of 64 triplets is divided into 8 equivalence classes with 8 triplets in each (Fig. 4). Each of these classes contains triplets with an identical dyadic numeration (that is the Hamming distance between dyadic numerations of triplets inside each equivalence class is equal to 0). One can see that this genomatrix [C A; G T]$^{(3)}$ is identical to the matrix of dyadic shifts on Figure 1 from the viewpoint of a disposition of members of the dyadic group (2).

$$\begin{vmatrix} C & A \\ G & T \end{vmatrix}^{(3)} = $$

| CCC 000 | CCA 001 | CAC 010 | CAA 011 | ACC 100 | ACA 101 | AAC 110 | AAA 111 |
|---|---|---|---|---|---|---|---|
| CCG 001 | CCT 000 | CAG 011 | CAT 010 | ACG 101 | ACT 100 | AAG 111 | AAT 110 |
| CGC 010 | CGA 011 | CTC 000 | CTA 001 | AGC 110 | AGA 111 | ATC 100 | ATA 101 |
| CGG 011 | CGT 010 | CTG 001 | CTT 000 | AGG 111 | AGT 110 | ATG 101 | ATT 100 |
| GCC 100 | GCA 101 | GAC 110 | GAA 111 | TCC 000 | TCA 001 | TAC 010 | TAA 011 |
| GCG 101 | GCT 100 | GAG 111 | GAT 110 | TCG 001 | TCT 000 | TAG 011 | TAT 010 |
| GGC 110 | GGA 111 | GTC 100 | GTA 101 | TGC 010 | TGA 011 | TTC 000 | TTA 001 |
| GGG 111 | GGT 110 | GTG 101 | GTT 100 | TGG 011 | TGT 010 | TTG 001 | TTT 000 |

Figure 4. The third Kronecker power of the genetic matrix [C A; G T] contains 64 triplets in a special order. Each triplet is a carrier of its binary numeration from dyadic group (2) if each pyrimidine is denoted by number "0" (C=T=0) and each purine is denoted by "1" (A=G=1).

This genetic matrix [C A; G T]$^{(3)}$ with dyadic numeration of triplets is one of useful tools to study hidden regularities in long nucleotide sequences. Let us apply this matrix to study repetitions of triplets in the whole human genome. Figure 5 shows quantities of repetitions of each triplet inside the whole human genome from the work [Perez, 2010]. This genome contains 2 843 411 612 (about three billion) triplets. A quantity of repetitions of a triplet inside nucleotide sequences is conditionally termed "a frequency of this triplet" or "a triplet frequency".

| triplet | triplet frequency | triplet | triplet frequency | triplet | triplet frequency | triplet | triplet frequency |
|---------|-------------------|---------|-------------------|---------|-------------------|---------|-------------------|
| AAA | 109143641 | CAA | 53776608 | GAA | 56018645 | TAA | 59167883 |
| AAC | 41380831 | CAC | 42634617 | GAC | 26820898 | TAC | 32272009 |
| AAG | 56701727 | CAG | 57544367 | GAG | 47821818 | TAG | 36718434 |
| AAT | 70880610 | CAT | 52236743 | GAT | 37990593 | TAT | 58718182 |
| ACA | 57234565 | CCA | 52352507 | GCA | 40907730 | TCA | 55697529 |
| ACC | 33024323 | CCC | 37290873 | GCC | 33788267 | TCC | 43850042 |
| ACG | 7117535 | CCG | 7815619 | GCG | 6744112 | TCG | 6265386 |
| ACT | 45731927 | CCT | 50494519 | GCT | 39746348 | TCT | 62964984 |
| AGA | 62837294 | CGA | 6251611 | GGA | 43853584 | TGA | 55709222 |
| AGC | 39724813 | CGC | 6737724 | GGC | 33774033 | TGC | 40949883 |
| AGG | 50430220 | CGG | 7815677 | GGG | 37333942 | TGG | 52453369 |
| AGT | 45794017 | CGT | 7137644 | GGT | 33071650 | TGT | 57468177 |
| ATA | 58649060 | CTA | 36671812 | GTA | 32292235 | TTA | 59263408 |
| ATC | 37952376 | CTC | 47838959 | GTC | 26866216 | TTC | 56120623 |
| ATG | 52222957 | CTG | 57598215 | GTG | 42755364 | TTG | 54004116 |
| ATT | 71001746 | CTT | 56828780 | GTT | 41557671 | TTT | 109591342 |

Figure 5. Quantities of repetitions of each triplet in the whole human genome (data are taken from the work [Perez, 2010])

One can see from this figure that quantities of repetitions of different triplets vary widely. For example, the triplet CGA is repeated 6 251 611 times and the triplet TTT is repeated 109 591 342 times (a difference of about 18 times). At first glance it seems that a set of these numbers is accidental. But the proposed division of a set of 64 triplets into the 8 equivalence classes reveals a hidden regularity in this pile of numbers.

By analogy with the dichotomous tree of a linguistic alphabet on Figure 3 let us present the set of 64 triplets as an alphabetical tree with a few levels of sub-alphabets (Figure 6). More precisely, in this genetic tree the first level of sub-alphabets consists of the following two sub-alphabets:

    1) 32 kinds of triplets with dyadic numerations 000, 001, 010 and 011 from Figure 4;
    2) 32 kinds of triplets with dyadic numerations 100, 101, 110 and 111 from Figure 4.

The second level of sub-alphabets in this genetic tree consists of the following four sub-alphabets:

    1) 16 kinds of triplets with dyadic numerations 000 and 010 from Figure 4;
    2) 16 kinds of triplets with dyadic numerations 001 and 011 from Figure 4;
    3) 16 kinds of triplets with dyadic numerations 100 and 110 from Figure 4;
    4) 16 kinds of triplets with dyadic numerations 101 and 111 from Figure 4.

The third level of sub-alphabets in this genetic tree consists of the following eight sub-alphabets:

    1) 8 kinds of triplets with dyadic numerations 000 from Figure 4;
    2) 8 kinds of triplets with dyadic numerations 010 from Figure 4;
    3) 8 kinds of triplets with dyadic numerations 001 from Figure 4;
    4) 8 kinds of triplets with dyadic numerations 011 from Figure 4;
    5) 8 kinds of triplets with dyadic numerations 100 from Figure 4;
    6) 8 kinds of triplets with dyadic numerations 110 from Figure 4;
    7) 8 kinds of triplets with dyadic numerations 101 from Figure 4;
    8) 8 kinds of triplets with dyadic numerations 111 from Figure 4.

Using data about frequency of different triplets from Figure 5, one can calculate the sum of triplet frequency in each sub-alphabet on all the three levels for the whole human genome. Figure 6 shows the interesting result of this action.

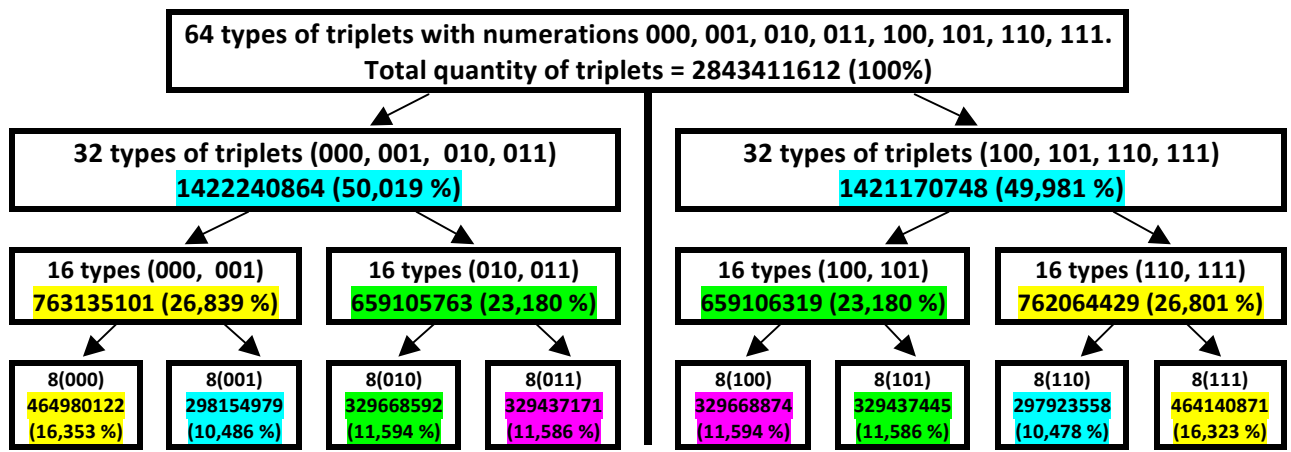

Figure 6. The first variant of the alphabetical dyadic tree of frequencies of triplets for the whole human genome on the base of the genetic matrix [C A; G T]$^{(3)}$ from Figure 4. Each cell on this tree shows a quantity of types of triplets, dyadic numerations of triplets from Figure 4, a quantity of repetitions of these triplets in the genome and the percentage of this quantity to the total number 2843411612 of all the triplets. This tree is symmetric (explanations in the text). Initial data about the genome are taken from [Perez, 2010].

As one can see from Figure 6, this alphabetical tree is symmetric because its left and right halves at each sub-alphabetical level contain paired sub-alphabets (or conjugate sub-alphabets), which are disposed mirror-symmetrical and which contain almost identical quantities of repetitions of the triplets. Each of pairs is marked by the same color for visibility. These paired sub-alphabets are not accidental and arbitrary because in all cases their dyadic numerations are inversion-symmetric: these numerations are transformed into each other by inversion of all the 0 into 1 and vice versa. For example, on the second level the sub-alphabet (001, 011) in the left half of the tree contains 659105763 repetitions of its triplets in the genome. Dyadic numerations 001, 011 in this sub-alphabet are transformed by the inversion 0→1 and 1→0 into dyadic numerations 110, 100 of the sub-alphabet in the right half with 659106319 repetitions of triplets (these conjugate sub-alphabets are marked by green on Figure 6). The ratio of these quantities 659106319 : 659105763 = 1,000000844; in other words the percentage difference between these conjugate sub-alphabets is equal to 0,00008%. The inversion-symmetric relations between dyadic numerations of members of paired sub-alphabets correspond to the fact that Hamming distances between these binary numerations are equal to maximum values.

| Dyadic numeration | Triplets with the dyadic numeration | Content of letters in sub-alphabet |
|---|---|---|
| 000 | CCC, CCT, CTC, CTT, TCC, TCT, TTC, TTT | 12 C, 12 T |
| 001 | CCA, CCG, CTA, CTG, TCA, TCG, TTA, TTG | 4 A, 8 C, 4 G, 8 T |
| 010 | CAC, CAT, CGC, CGT, TAC, TAT, TGC, TGT | 4 A, 8 C, 4 G, 8 T |
| 011 | CAA, CAG, CGA, CGG, TAA, TAG, TGA, TGG | 8 A, 4 C, 8 G, 4 T |
| 100 | ACC, ACT, ATC, ATT, GCC, GCT, GTC, GTT | 4 A, 8 C, 4 G, 8 T |
| 101 | ACA, ACG, ATA, ATG, GCA, GCG, GTA, GTG | 8 A, 4 C, 8 G, 4 T |
| 110 | AAC, AAT, AGC, AGT, GAC, GAT, GGC, GGT | 8 A, 4 C, 8 G, 4 T |
| 111 | AAA, AAG, AGA, AGG, GAA, GAG, GGA, GGG | 12 A, 12 G |

Figure 7. The content of nitrogenous bases A, C, G and T in each of the sub-alphabets of the third level of the alphabetic tree on Figure 6.

Figure 7 shows a content of nitrogenous bases A, C, G and T in each of sub-alphabets of the third level of the alphabetic tree on Figure 6. One can see that total quantities of the complementary bases A and T (and also C and G) are identical in both paired sub-alphabets of this basic level. For example, each of the paired sub-alphabets with dyadic numerations 011 and 100 contains the total quantity 12 of the letters A and T and also the total quantity 12 of the letters C and G. But in accordance with Chargaff's second parity rule, quantities of the complementary letters A and T (and also C and G) are approximately equal to each other: in the whole human genome, adenine A is repeated 2519482582 times, thymine T – 2523644402 times, cytosine C -1743079028 times, guanine G – 1744028824 times [Perez, 2010, table 8]. One can conclude that the quantitative features of the dyadic tree (Figure 6) correspond to the Chargaff's second parity rule.

| № | 000 | 001 | 010 | 011 | 100 | 101 | 110 | 111 |
|---|-----|-----|-----|-----|-----|-----|-----|-----|
| 1 | 15973 | 11636 | 10855 | 11421 | 11532 | 10750 | 11269 | 16564 |
| 2 | 16841 | 11730 | 11079 | 11424 | 11680 | 10528 | 11433 | 15285 |
| 3 | 16494 | 11578 | 10465 | 11432 | 11410 | 10691 | 11514 | 16416 |
| 4 | 16031 | 11611 | 10895 | 11436 | 11404 | 10781 | 11469 | 16373 |
| 5 | 16638 | 11740 | 11414 | 11313 | 11624 | 11007 | 11406 | 14858 |
| 6 | 16551 | 11577 | 11339 | 11363 | 11544 | 11092 | 11493 | 15041 |

figure 8. Quantities of triplets in sub-alphabets of the third level (Figure 6) with appropriate dyadic numerations (000, 001, …, 111) in the following long nucleotide sequences with 100000 triplets in each: 1) Homo sapiens genomic DNA, chromosome 1q22-q23, CD1 region, section 3/4, GenBank: AP002534.1; 2) Homo sapiens genomic DNA, chromosome 6q21, anti-oncogene region, section 1/4, GenBank: AP002528.1; 3) Homo sapiens genomic DNA, chromosome 1q22-q23, CD1 region, section 1/4, GenBank: AP002532.1; 4) Homo sapiens genomic DNA, chromosome 1q22-q23, CD1 region, section 2/4, GenBank: AP002533.1; 5) Homo sapiens genomic DNA, chromosome 6q21, anti-oncogene region, section 2/4, GenBank: AP002529.1; 6) Homo sapiens genomic DNA, chromosome 6q21, anti-oncogene region, section 3/4, GenBank: AP002530.2. Initial data are taken from http://www.ncbi.nlm.nih.gov/sites/entrez.

Figure 8 shows some results of a spot check of symmetric properties of the alphabetic trees for a series of long sequences of nucleotides. This check confirms symmetric properties of the alphabetic trees for cases of long nucleotide sequences. For this check the author has used a possibility of a choice of nucleotide sequences of arbitrary lengths on the site of http://www.ncbi.nlm.nih.gov/sites/entrez. On this site the author has asked the search of sequences of nucleotides with lengths between 300000 and 300001 by means of the following typical instruction: 300000:300001[SLEN]. Figure 8 shows results for cases of the first five nucleotide sequences there with 300000 nucleotides (or 100000 triplets) in each. These data on Figure 8 are enough to construct alphabetic trees for these nucleotide sequences by analogy with Figure 6.

4. **The second variant of dyadic numerations of triplets and dyadic trees of triplet frequencies in long nucleotide sequences**

This section describes another Kronecker family of the genetic matrices based on the kernel (2*2)-matrix [C T; G A] (Figure 9). In accordance with the second sub-alphabet on Figure 2 one can denote each amino by binary number "0" (C=A=0) and each keto by binary number "1" (G=T=1). In such case each triplet obtains its corresponding binary numeration depending on its three nitrogenous bases. For example the triplet CAG obtains the binary numeration 001. The whole set of 64 triplets is divided into 8 equivalence classes with 8 triplets in each (Fig. 9). Each of these classes contains triplets with an identical dyadic numeration (that is the Hamming distance between dyadic numerations of triplets inside each equivalence class is equal to 0). One can see that this genomatrix [C T; G A]^(3) is identical to the matrix of dyadic shifts on Figure 1 from the viewpoint of a disposition of members of the dyadic group (2).

$$\begin{vmatrix} C & T \\ G & A \end{vmatrix}^{(3)} =$$

| CCC 000 | CCT 001 | CTC 010 | CTT 011 | TCC 100 | TCT 101 | TTC 110 | TTT 111 |
|---|---|---|---|---|---|---|---|
| CCG 001 | CCA 000 | CTG 011 | CTA 010 | TCG 101 | TCA 100 | TTG 111 | TTA 110 |
| CGC 010 | CGT 011 | CAC 000 | CAT 001 | TGC 110 | TGT 111 | TAC 100 | TAT 101 |
| CGG 011 | CGA 010 | CAG 001 | CAA 000 | TGG 111 | TGA 110 | TAG 101 | TAA 100 |
| GCC 100 | GCT 101 | GTC 110 | GTT 111 | ACC 000 | ACT 001 | ATC 010 | ATT 011 |
| GCG 101 | GCA 100 | GTG 111 | GTA 110 | ACG 001 | ACA 000 | ATG 011 | ATA 010 |
| GGC 110 | GGT 111 | GAC 100 | GAT 101 | AGC 010 | AGT 011 | AAC 000 | AAT 001 |
| GGG 111 | GGA 110 | GAG 101 | GAA 100 | AGG 011 | AGA 010 | AAG 001 | AAA 000 |

Figure 9. The third Kronecker power of the genetic matrix [C T; G A] contains 64 triplets in a special order. Each triplet is a carrier of its binary numeration from dyadic group (2) if each amino is denoted by number "0" (C=A=0) and each keto is denoted by "1" (G=T=1) in accordance with the second sub-alphabet (Figure 2)

| № | | | | | | | | |
|---|---|---|---|---|---|---|---|---|
| 0 | 64 types of triplets with numerations 000, 001, 010, 011, 100, 101, 110, 111. Total quantity of triplets = 2843411612 (100%) | | | | | | | |
| 1 | 32 вида (000, 001, 010, 011) 1420853917 (49,970 %) | | | | 32 вида (100, 101, 110, 111) 1422557695 (50,030 %) | | | |
| 2 | 16 видов (000, 001) 775361012 (27,269 %) | | 16 видов (010, 011) 645492905 (22,701 %) | | 16 видов (100, 101) 645492860 (22,701 %) | | 16 видов (110, 111) 777064835 (27,329 %) | |
| 3 | 8(000) 426837965 (15,011%) | 8(001) 348523047 (12,257%) | 8(010) 296663649 (10,433 %) | 8(011) 348829256 (12,268 %) | 8(100) 348523003 (12,257%) | 8(101) 296969857 (10,444%) | 8(110) 348829204 (12,268%) | 8(111) 428235631 (15,061%) |

Figure 10. The second variant of the alphabetical dyadic tree of frequencies of triplets for the whole human genome on the base of the genetic matrix [C T; G A]^(3) from Figure 9. The left column shows numerations of levels of sub-alphabets. Each cell on this tree shows a quantity of types of triplets, dyadic numerations of triplets from Figure 9, a quantity of repetitions of these triplets in the genome and the percentage of this quantity to the total number 2843411612 of all the triplets. This tree is symmetric (explanations in the text). Initial data are taken from [Perez, 2010].

As one can see from Figure 10, this alphabetical tree is also symmetric because its left and right halves at each sub-alphabetical level contain paired sub-alphabets, which are disposed mirror-symmetrical and which contain almost identical numbers. Each of pairs is marked by the same color for visibility. These paired sub-alphabets are not accidental and arbitrary because in all cases their dyadic numerations are inversion-symmetric: these numerations are transformed into each other by inversion of all the 0 into 1 and vice versa. The inversion-symmetric relations between dyadic numerations of members of paired sub-alphabets correspond to the fact that Hamming distances between these binary numerations are equal to maximum values.

| Dyadic numeration | Triplets with the dyadic numeration | Content of letters in sub-alphabet |
|---|---|---|
| 000 | CCC CCA CAC CAA ACC ACA AAC AAA | 12 A, 12 C |
| 001 | CCT CCG CAT CAG ACT ACG AAT AAG | 8 A, 8 C, 4 G, 4 T |
| 010 | CTC, CTA, CGC, CGA ATC ATA AGC AGA | 8 A, 8 C, 4 G, 4 T |
| 011 | CTT CTG CGT CGG ATT ATG AGT AGG | 4 A, 4 C, 8 G, 8 T |
| 100 | TCC, TCA, TAC, TAA, GCC GCA GAC GAA | 8 A, 8 C, 4 G, 4 T |
| 101 | TCT TCG TAT TAG GTC GTA GGC GGA | 4 A, 4 C, 8 G, 8 T |
| 110 | TTC TTA TGC TGA GTC GTA GGC GGA | 4 A, 4 C, 8 G, 8 T |
| 111 | TTT TTG TGT TGG GTT GTG GGT GGG | 12 G, 12 T |

Figure 11. The content of nitrogenous bases A, C, G and T in each of the sub-alphabets of the third level of the alphabetic tree on Figure 10.

Figure 11 shows a content of nitrogenous bases A, C, G and T in each of sub-alphabets of the third level of the alphabetic tree on Figure 10. One can see that total quantities of the complementary bases A and T (and also C and G) are identical in both paired sub-alphabets of this basic level. For example, each of the paired sub-alphabets with dyadic numerations 011 and 100 contains the total quantity 12 of the complementary letters A and T and also the total quantity 12 of the complementary letters C and G. But in accordance with Chargaff's second parity rule, quantities of the complementary letters A and T (and also C and G) are approximately equal to each other. One can conclude that the quantitative features of the alphabetic tree (Figure 10) correspond also to the Chargaff's second parity rule.

A spot check of the alphabetic trees for a series of long sequences of nucleotides confirms symmetric properties of the alphabetic trees for cases of long nucleotide sequences by close analogy with the results on Figure 8.

Let us turn now to the third binary sub-alphabet (Figure 4) and consider the genetic matrix [C T; A G]$^{(3)}$ with 64 triplets inside it. In accordance with the third binary sub-alphabet one can denote the complementary letters C and G by binary number 0 (C=G=0) and the complementary letters A and T by binary number 1 (A=T=1). In this case each of triplets possesses its own dyadic numeration, and the matrix [C T; A G]$^{(3)}$ appears as the dyadic-shift matrix from the viewpoint of the disposition of 64 dyadic numerations of triplets inside it. But the alphabetic tree on the basis of this matrix [C T; A G]$^{(3)}$ does not have symmetric properties, in contrast to the cases of two matrices [C A; G T]$^{(3)}$ and [C T; G A]$^{(3)}$ discussed above (Figures 4 and 9). Here it should be noted the essential difference of the third sub-alphabet from the first two sub-alphabets on Figure 2. Really, the first two binary sub-alphabets rely upon the intrinsic molecular characteristics of the nitrogenous bases: purine or pyrimidine, and amino or keto. But the third binary sub-alphabet relies only upon the external characteristics of these molecules: two or three hydrogen bonds participate in their complementary connection.

5. **The genetic code and linguistics**

Above we described some new analogies in the form of dichotomous trees between alphabetic systems of linguistics and the genetic code. It gives the opportunity to recall the known theme of the relationship of linguistics to the genetic code which was described, for example, in works [Petoukhov, 2003-2004, 2008; Petoukhov, He, 2010]. Impressive discoveries in the field of the genetic code have been described by its researchers using the terminology borrowed from linguistics and the theory of communications. As experts in molecular genetics remark, "*the more we understand laws of coding of the genetic information, the more strongly we are surprised by their similarity to principles of linguistics of human and computer languages*" [Ratner, 2002, p. 203]. Linguistics is one of the significant examples of existence and importance of ensembles of binary oppositions in information physiology.

Leading experts on structural linguistics have believed for a long time that languages of human dialogue were formed not from an empty place, but they are continuation of genetic language or, somehow, are closely connected with it, confirming the idea of information commonality of organisms. Analogies between systems of genetic and linguistic information are contents of a wide and important scientific sphere, which can be illustrated here in short only. We reproduce below some thematic thoughts by R. Jakobson [1985, 1987, 1999], who is one of the most famous experts and the author of a deep theory of binary oppositions in linguistics. Jakobson and others are holding the same views that we possess a language, which is as old as life and which is the most alive among all languages. Among all systems of information transfer, the genetic code and linguistic codes only are based on the use of discrete components, which in itself makes no sense, but serve for the construction of the minimum units, which make sense. In both cases of the genetic language and of a linguistic language, we deal with separate units which, taken in itself, have no sense, but they get a sense after their special grouping. (By the way, one can note here that matrix genetics deals with matrix forms of groupings of elements of genetic language successfully). A similarity between both information systems is not exhausted by this fact at all. According to Jakobson, all relations among linguistic phonemes are decomposed into a series of binary oppositions of elementary differential attributes (or traits). By analogy the set of the four letters of the genetic alphabet contains the three binary sub-alphabets (Figure 2) which provides dyadic numerations of columns and rows of the genetic matrices together with genetic multiplets (triplets, duplets, etc.). As Jakobson stated, the genetic code system is the basic simulator, which underlines all verbal codes of human languages. "*The heredity in itself is the fundamental form of communications ... Perhaps, the bases of language structures, which are imposed on molecular communications, have been constructed by its structural principles directly*" [Jakobson, 1985, p. 396]. These questions have arisen to Jakobson as a consequence of its long-term researches of connections between linguistics, biology and physics. Such connections were considered at a united seminar of physicists and linguists, which was organized by Niels Bohr and Roman Jakobson jointly at the Massachusetts Institute of Technology.

"*Jakobson reveals distinctly a binary opposition of sound attributes as underlying each system of phonemes... The subject of phonology has changed by him: the phonology considered phonemes (as the main subject) earlier, but now Yakobson has offered that distinctive attributes should be considered as "quantums" (or elementary units of language)… . Jakobson was interested especially in the general analogies of language structures with the genetic code, and he considered these analogies as indubitable*" [Ivanov, 1985]. One can remind also of the title of the monograph "On the Yin and Yang nature of language" [Baily, 1982], which is characteristic for the theme of binary oppositions in linguistics.

Similar questions about a connection of linguistics with the genetic code excite many researchers. In addition many researchers perceive a linguistic language as a living organism. The book "Linguistic genetics" [Makovskiy, 1992] states: "*The opinion about language as about a living organism, which is submitted to the laws of a nature, ascends to a deep antiquity ... Research of a nature, of disposition and of reasons of isomorphism between genetic and linguistic regularities is one of the most important fundamental problems for linguistics of our time*".

One of the interesting questions is the existence of fractal images in linguistic and genetic texts. A number of publications are devoted to fractal features of linguistic and genetic texts [Gariaev, 1994; Jeffry, 1990; Yam, 1995, etc]. Interesting data about fractal approaches in genetics are presented in [Pellionisz et al, 2011; http://www.junkdna.com/the_genome_is_fractal.html]). Researches in this direction proceed all over the world.

We believe that achievements of matrix genetics and its mathematical notions and tools will be useful for revealing deep connections between genetic and linguistic languages. This matrix-genetic approach is capable of enriching its own arsenal of structural linguistics as a roughly developing science and to clear a problem of the unified bases of biological languages. It can be applied to researches on evolutionary linguistics, the analysis and synthesis of poetic forms, etc.

## 6. Dual numbers and their extensions in the field of matrix genetics

In previous sections we considered Kronecker families of the genetic matrices [C A; G T] (Figure 4) and [C T; G A] (Figure 9) and the appropriate alphabetic trees (Figures 6 and 10). In this section we continue a consideration of such matrices to show their connections with the matrix representation of dual numbers [b 0; d b] and briefly with Fibonacci matrices (Section 7 contains more details about Fibonacci matrices). It is important to construct mathematical models, first of all, of the following inherited phenomena in living matter: 1) spiral formations in a great number of biological structures; 2) phyllotaxis laws which are connected with Fibonacci numbers. Results of this section about different numeric representations of symbolic genetic matrices on the base of molecular peculiarities of their genetic elements give additional materials to the principle "multi-lingual genetics" from the Section 1: "a piece of a genetic text - various programs for its reading" depending on molecular traits which are taken into account when reading genetic sequences.

Here one can remind about the pioneer work [Konopelchenko & Rumer, 1975a, 1975b] in the field of matrix genetics. This work was published twice: in the most prestigious scientific journal of the USSR and in a form of the preprint in English. This work presented the 4-letter genetic alphabet C, A, G, U/T in a form of a (2x2)-matrix [C G; U A]. After this it considered the second Kronecker power of this alphabetic matrix, which generated a (4x4)-matrix of 16 genetic duplets for investigation of symmetrical and other properties of this ensemble of genetic components. A special numeric representation of this (4x4)-genomatrix was also considered.

Let us note that G.Rumer was the main co-author of this pioneer article and he was a prominent Russian scientist in the field of theory of symmetry. Under a personal recommendation by A.Einstein and P. Ehrenfest he obtained a Lorentz's grant and worked as an assistant of M.Born in Gottingen in the period of 1929-1932 years. In the co-authorship with H.Weyl, V.Heitler and E.Teller, Rumer has created bases of quantum chemistry. Rumer believed that properties of symmetry play an essential role in phenomenology of the genetic code. His works on classification of codons in the genetic code, based on principles of symmetry and linguistic reasons, have obtained a benevolent response by F.Crick (see more details about G.Rumer in [Petoukhov, He, 2010, Chapter 1; Ginzburg, Mihailov & Pokrovskiy, 2001; web-site http://www.nsu.ru/assoz/rumer/]).

Our own research on matrix genetics continued the line of research presented in this brief publication [Konopelchenko & Rumer, 1975a, 1975b], contributing in this area, for example, the following new approaches and results: 1) the third and higher Kronecker degrees of (2*2)-matrices are considered; 2) taking into account the molecular characteristics of genetic elements, the author constructs new numerical representations of genetic matrices of Kronecker families; 3) these numerical representations of genetic matrices are analyzed from the viewpoint of their connections with the hypercomplex numbers, including Hamilton's quaternions, Cockle's split-quaternions and some others, which play essential role in mathematical natural sciences; 4) close relationships of the genetic system with dyadic shifts and modulo-2 addition are revealed; 5) close relationships of the genetic matrices with Hadamard matrices and with mathematical tools of noise-immune coding are revealed; 6) the principle of "multi-lingual genetics" is put forward, etc.

Below we will pay a special attention to the genetic matrix [C G; T A], which was considered in the pioneer work [Konopelchenko & Rumer, 1975a, 1975b] (we replaced the letter U by the letter T in this matrix). The genomatrix of triplets [C G; T A]$^{(3)}$ is the transposed analogue of the genomatrix [C T; G A]$^{(3)}$, which was considered above (Figure 9). Both of these genomatrices [C G; T A]$^{(3)}$ and [C T; G A]$^{(3)}$ define the same alphabetical dyadic tree (Figure 10) because of symmetric dispositions of members of the dyadic group 000, 001, …, 111 inside the (8*8)-matrix of dyadic shifts. In order to proceed to numerical representations of these matrices, we first consider the molecular features of nitrogenous bases A, C, G, T/U.

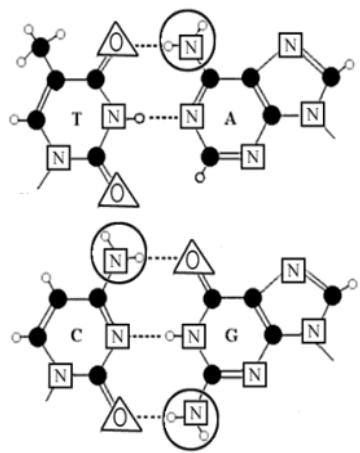

Figure 12. The complementary pairs of the 4 nitrogenous bases in DNA: A-T (adenine and thymine), C-G (cytosine and guanine). Hydrogen bonds in these pairs are shown by dotted lines. Black circles are atoms of carbon; small white circles are atoms of hydrogen; squares with the letter N are atoms of nitrogen; triangles with the letter O are atoms of oxygen. Amides (or amino-groups) $NH_2$ are marked by big circles (from [Petoukhov, He, 2010]).

Figure 12 shows the set of 4 nitrogenous bases in DNA: A, T, G and C. Molecular peculiarities of these 4 poly-atomic objects present two variants of the division of this set into the sub-set of three letters and a sub-set of one letter:
1) The first variant is based on the fact that the three nitrogenous bases A, C and G have one amides (amino-group) $NH_2$, but the fourth basis T has not it. The amino-group $NH_2$ has important functional meaning [Petoukhov, He, 2010, p.119]. One should note else that letter T is different additionally from the other three letters A, C and G by the fact that it is replaced by another letter U for some unknown reason under the transition from DNA to RNA. Thus, taking into account the special status of the letter T, the reasoned division of the four-letter set A, C, G and T into the 3-letter sub-alphabet A, C and G, and into the 1-letter sub-alphabet T exists (Figure 13, left).
2) The second variant is based on the fact that the three nitrogenous bases C, G and T have oxygen (Figure 12) but the fourth basis A has not it. It is well known that oxygen plays an important biological role. Thus, taking into account the special status of the letter A, the reasoned division of the four-letter set A, C, G and T into the 3-letters sub-alphabet C, G and T, and into the 1-letter sub-alphabet A (Figure 13, right) exists.

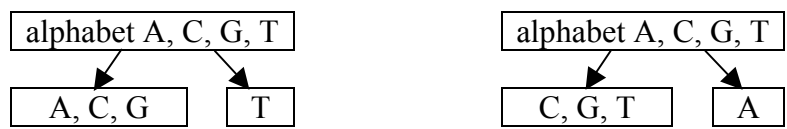

Figure 13. Two variants of the reasoned divisions of the 4-letter set of nitrogenous letters A, C, G and T into the 3-letter sub-alphabet and the 1-letter sub-alphabet (explanation in text).

In information technologies it is usual way to denote binary-oppositional objects by means of binary numbers "1" and "0" (or sometimes by means of numbers "+1" and "-1"). By analogy one can suppose that both these pairs of binary-oppositional symbols ("1" and "0"; or

"+1" and "-1") together with mathematical operations with them are utilized in a computer of organism as well. In this case, the same binary-oppositional traits of molecular-genetic systems can be expressed in genomatrices by means of the pair of binary elements "+1" and "-1" or by means of the pair of binary elements "0" and "1". It increases possibilities of bioinformation technologies significantly. The possibility of a similar different interpretation of the same material depending on surrounding conditions is known in genetics: for example, the trait of alopecia is dominating for men, but this trait is recessive for women. It means that this trait depends on the internal environment of an organism (see more details in [Petoukhov, 2008a; Petoukhov, He, 2010]).

Now for the first alphabet (Figure 13, left) let us denote all the members of the 3-letter sub-alphabet by means of "+1" (A=C=G=1) and the letter T from the second sub-alphabet - by means of "0" (T=0). In this case the symbolic genomatrices matrices [C A; G T], [C T; G A] and [C T; A G] are transformed into their numeric representations on Figure 14.

$$\begin{bmatrix} C & A \\ G & T \end{bmatrix} \rightarrow F = \begin{bmatrix} 1 & 1 \\ 1 & 0 \end{bmatrix}; \quad \begin{bmatrix} C & G \\ T & A \end{bmatrix} \rightarrow D_1 = \begin{bmatrix} 1 & 1 \\ 0 & 1 \end{bmatrix}; \quad \begin{bmatrix} C & A \\ T & G \end{bmatrix} \rightarrow D_1 = \begin{bmatrix} 1 & 1 \\ 0 & 1 \end{bmatrix}$$

Figure 14. Numeric representations of the genetic matrices [C A; G T], [C G; T A] and [C A; T G] for the case of the first alphabet (Figure 13, left) when A=C=G=1 and T=0. The matrix F is the Fibonacci matrix (see Section 7), and the matrix $D_1$ is the matrix representation of the dual number x+y*w, whose coordinates are equal to 1 (here $w^2=0$, x=y=1).

In the case of the second alphabet (Figure 13, right) one can again denote all the members of the 3-letter sub-alphabet by means of "+1" (C=G=T=1) and the letter A from the second sub-alphabet - by means of "0" (A=0). In this case the symbolic genomatrices matrices [C A; G T], [C G; T A] and [C T; A G] are transformed into their numeric presentations on Figure 15.

$$\begin{bmatrix} C & A \\ G & T \end{bmatrix} \rightarrow D_2 = \begin{bmatrix} 1 & 0 \\ 1 & 1 \end{bmatrix}; \quad \begin{bmatrix} C & G \\ T & A \end{bmatrix} \rightarrow F = \begin{bmatrix} 1 & 1 \\ 1 & 0 \end{bmatrix}; \quad \begin{bmatrix} C & T \\ A & G \end{bmatrix} \rightarrow D_1 = \begin{bmatrix} 1 & 1 \\ 0 & 1 \end{bmatrix}$$

Figure 15. Numeric presentations of the genetic matrices [C A; G T], [C G; T A] and [C T; A G] for the case of the second alphabet (Figure 13, right) when C=G=T=1 and A=0. The matrix F is the Fibonacci matrix, and the matrices $D_1$ and $D_2$ are the matrix presentations of the dual number x+y*w, whose coordinates are equal to 1 (here $w^2=0$, x=y=1).

In cases of the both alphabets (Figure 13, left and right) we receive on Figure 14 and 15 the types of numeric matrices which are well-known in mathematics: the Fibonacci matrix F (see Section 7) and the matrix presentations $D_1$ and $D_2$ of the dual number x+y*w (here $w^2=0$) whose coordinates are equal to 1 (x=y=1). Figure 16 shows decompositions of known matrix representations of dual numbers x+y*w where x, y are real numbers and $w^2=0$, w≠1 (see the web-site on dual numbers in mathematics and its applications http://en.wikipedia.org/wiki/Dual_number). One can see from Figures 14 and 15 that the same genomatrix [C G; T A] (or [C A; G T]) can be read as the Fibonacci matrix or the dual number with unit coordinates depending on a molecular designation: 1) A=C=G=1, T=0 or 2) T=C=G=1, A=0.

$$\begin{bmatrix} x & y \\ 0 & x \end{bmatrix} = x*\begin{bmatrix} 1 & 0 \\ 0 & 1 \end{bmatrix} + y*\begin{bmatrix} 0 & 1 \\ 0 & 0 \end{bmatrix}; \quad \begin{bmatrix} x & 0 \\ y & x \end{bmatrix} = x*\begin{bmatrix} 1 & 0 \\ 0 & 1 \end{bmatrix} + y*\begin{bmatrix} 0 & 0 \\ 1 & 0 \end{bmatrix}; \quad \begin{bmatrix} 1 & w \\ w & 0 \end{bmatrix} \begin{matrix} 1 & w \\ 1 & w \end{matrix}$$

Figure 16. Decompositions of matrix representations of dual numbers x+y*w. The multiplication table of the 2-dimensional algebra of dual numbers is shown on the right. Here w=[0 1; 0 0] or w=[0 0; 1 0].

Below the case of the genomatrix [C G; T A] from the work [Konopelchenko & Rumer, 1975a, 1975b] is considered attentively. Other cases of genomatrices ([C A; G T], [C T; A G], etc.) can be considered by analogy.

Figure 17 shows the genomatrix [C G; T A]$^{(2)}$ and its numeric representation in the molecular reasonable case of the denotation C=G=A=1 and T=0 in the three cases: 1) both positions of duplets are taken into account, and each duplet is represented as a result of multiplication of numbers 0 and 1 in accordance with its letters (for example the duplet CT is represented by zero because of 1*0 is equal to zero); 2) only the first position of duplets is taken into account; 3) only the second position of duplets is taken into account. The dyadic-shift decomposition of each numeric representation of matrices on Figure 17 gives a set of four sparse matrices $d_0, d_1, d_2, d_3$ (this set is individual for each of the three cases) which is closed in relation to multiplication; its multiplication table corresponds to a 4-dimensional algebra which is an extension of the 2-dimensional algebra of dual numbers with the multiplication table on Figure 16, on the right.

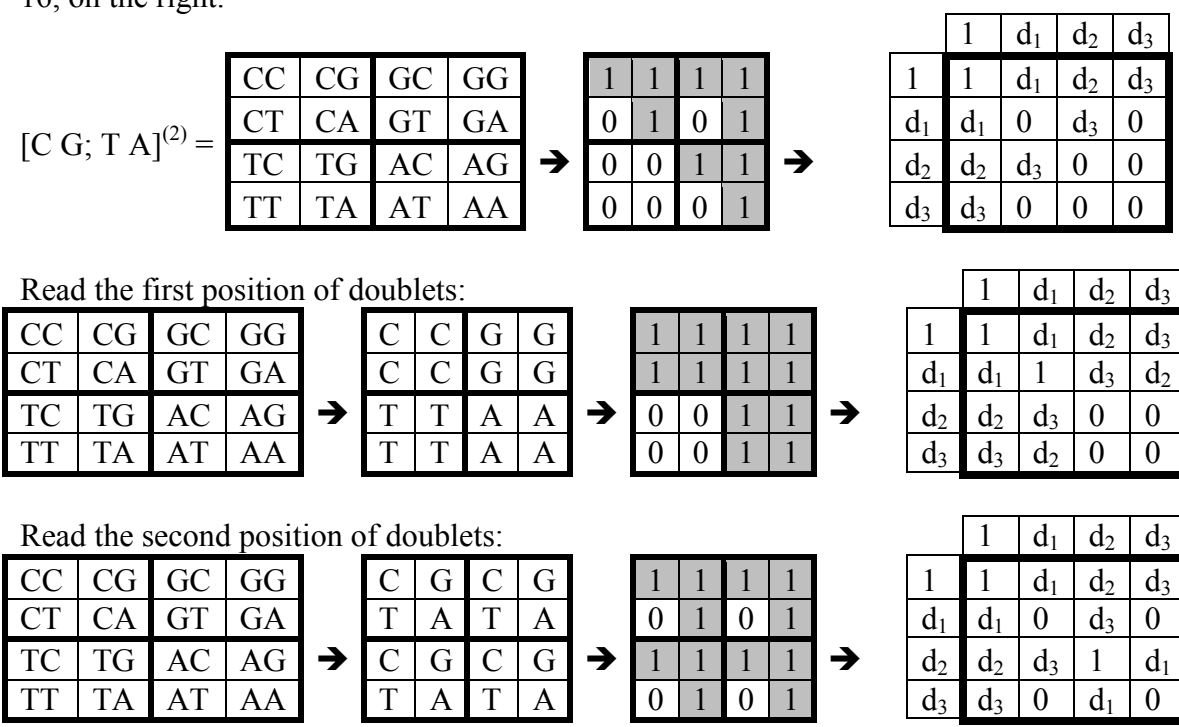

Figure 17. The genomatrix [C G; T A]$^{(2)}$ and the following three cases of its numeric representation under the molecular reasonable denotation C=G=A=1 and T=0: 1) the upper row: both positions of duplets are taken into account, and each duplet is represented as a result of multiplication of numbers 0 and 1 in accordance with its letters; 2) the middle row: only the first position of duplets is taken into account; 3) the lower row: only the second position of duplets is taken into account. In each row, the right column contains a multiplication table for the appropriate set of four sparse matrices $d_0, d_1, d_2, d_3$ of the dyadic-shift decomposition of the relevant numeric matrix (in each row $d_0$ is the unit matrix). Each of these three multiplication tables coincides with a multiplication table of a 4-dimensional algebra, which is an extension of the 2-dimensional algebra of dual numbers (Figure 16, on the right). Black (white) cells contain entries 1 (0).

Figure 18 shows the genomatrix [C G; T A]$^{(3)}$ of 64 triplets and its numeric representation under the same denotation C=G=A=1 and T=0 in the four cases: 1) all positions of triplets are taken into account, and each triplet is represented as a result of multiplication of numbers 0 and 1 in accordance with its letters (for example the triplet CTA is represented by zero because of 1*0*1 is equal to zero); 2) only the first position of triplets is taken into account; 3) only the second position of triplets is taken into account; 4) only the third position of triplets is taken into account. The dyadic-shift decomposition of each numeric representation of matrices on Figure 18 gives a set of eight sparse matrices $d_0, d_1, d_2, d_3, d_4, d_5, d_6, d_7$ (this set is individual for each of the four cases) which is closed in relation to multiplication; its multiplication table corresponds to a 8-dimensional algebra which is an extension of the 2-dimensional algebra of dual numbers with the multiplication table on Figure 16, on the right.

$[C\ G;\ T\ A]^{(3)} =$

| CCC | CCG | CGC | CGG | GCC | GCG | GGC | GGG |
|-----|-----|-----|-----|-----|-----|-----|-----|
| CCT | CCA | CGT | CGA | GCT | GCA | GGT | GGA |
| CTC | CTG | CAC | CAG | GTC | GTG | GAC | GAG |
| CTT | CTA | CAT | CAA | GTT | GTA | GAT | GAA |
| TCC | TCG | TGC | TGG | ACC | ACG | AGC | AGG |
| TCT | TCA | TGT | TGA | ACT | ACA | AGT | AGA |
| TTC | TTG | TAC | TAG | ATC | ATG | AAC | AAG |
| TTT | TTA | TAT | TAA | ATT | ATA | AAT | AAA |

→

| 1 | 1 | 1 | 1 | 1 | 1 | 1 | 1 |
|---|---|---|---|---|---|---|---|
| 0 | 1 | 0 | 1 | 0 | 1 | 0 | 1 |
| 0 | 0 | 1 | 1 | 0 | 0 | 1 | 1 |
| 0 | 0 | 0 | 1 | 0 | 0 | 0 | 1 |
| 0 | 0 | 0 | 0 | 1 | 1 | 1 | 1 |
| 0 | 0 | 0 | 0 | 0 | 1 | 0 | 1 |
| 0 | 0 | 0 | 0 | 0 | 0 | 1 | 1 |
| 0 | 0 | 0 | 0 | 0 | 0 | 0 | 1 |

→

|       | 1     | $d_1$ | $d_2$ | $d_3$ | $d_4$ | $d_5$ | $d_6$ | $d_7$ |
|-------|-------|-------|-------|-------|-------|-------|-------|-------|
| 1     | 1     | $d_1$ | $d_2$ | $d_3$ | $d_4$ | $d_5$ | $d_6$ | $d_7$ |
| $d_1$ | $d_1$ | 0     | $d_3$ | 0     | $d_5$ | 0     | $d_7$ | 0     |
| $d_2$ | $d_2$ | $d_3$ | 0     | 0     | $d_6$ | $d_7$ | 0     | 0     |
| $d_3$ | $d_3$ | 0     | 0     | 0     | $d_7$ | 0     | 0     | 0     |
| $d_4$ | $d_4$ | $d_5$ | $d_6$ | $d_7$ | 0     | 0     | 0     | 0     |
| $d_5$ | $d_5$ | 0     | $d_7$ | 0     | 0     | 0     | 0     | 0     |
| $d_6$ | $d_6$ | $d_7$ | 0     | 0     | 0     | 0     | 0     | 0     |
| $d_7$ | $d_7$ | 0     | 0     | 0     | 0     | 0     | 0     | 0     |

Read the first position of triplets:

| C | C | C | C | G | G | G | G |
|---|---|---|---|---|---|---|---|
| C | C | C | C | G | G | G | G |
| C | C | C | C | G | G | G | G |
| C | C | C | C | G | G | G | G |
| T | T | T | T | A | A | A | A |
| T | T | T | T | A | A | A | A |
| T | T | T | T | A | A | A | A |
| T | T | T | T | A | A | A | A |

→

| 1 | 1 | 1 | 1 | 1 | 1 | 1 | 1 |
|---|---|---|---|---|---|---|---|
| 1 | 1 | 1 | 1 | 1 | 1 | 1 | 1 |
| 1 | 1 | 1 | 1 | 1 | 1 | 1 | 1 |
| 1 | 1 | 1 | 1 | 1 | 1 | 1 | 1 |
| 0 | 0 | 0 | 0 | 1 | 1 | 1 | 1 |
| 0 | 0 | 0 | 0 | 1 | 1 | 1 | 1 |
| 0 | 0 | 0 | 0 | 1 | 1 | 1 | 1 |
| 0 | 0 | 0 | 0 | 1 | 1 | 1 | 1 |

→

|       | 1     | $d_1$ | $d_2$ | $d_3$ | $d_4$ | $d_5$ | $d_6$ | $d_7$ |
|-------|-------|-------|-------|-------|-------|-------|-------|-------|
| 1     | 1     | $d_1$ | $d_2$ | $d_3$ | $d_4$ | $d_5$ | $d_6$ | $d_7$ |
| $d_1$ | $d_1$ | 1     | $d_3$ | $d_2$ | $d_5$ | $d_4$ | $d_7$ | $d_6$ |
| $d_2$ | $d_2$ | $d_3$ | 1     | $d_1$ | $d_6$ | $d_7$ | $d_4$ | $d_5$ |
| $d_3$ | $d_3$ | $d_2$ | $d_1$ | 1     | $d_7$ | $d_6$ | $d_5$ | $d_4$ |
| $d_4$ | $d_4$ | $d_5$ | $d_6$ | $d_7$ | 0     | 0     | 0     | 0     |
| $d_5$ | $d_5$ | $d_4$ | $d_7$ | $d_6$ | 0     | 0     | 0     | 0     |
| $d_6$ | $d_6$ | $d_7$ | $d_4$ | $d_5$ | 0     | 0     | 0     | 0     |
| $d_7$ | $d_7$ | $d_6$ | $d_5$ | $d_4$ | 0     | 0     | 0     | 0     |

Read the second position of triplets:

| C | C | G | G | C | C | G | G |
|---|---|---|---|---|---|---|---|
| C | C | G | G | C | C | G | G |
| T | T | A | A | T | T | A | A |
| T | T | A | A | T | T | A | A |
| C | C | G | G | C | C | G | G |
| C | C | G | G | C | C | G | G |
| T | T | A | A | T | T | A | A |
| T | T | A | A | T | T | A | A |

→

| 1 | 1 | 1 | 1 | 1 | 1 | 1 | 1 |
|---|---|---|---|---|---|---|---|
| 1 | 1 | 1 | 1 | 1 | 1 | 1 | 1 |
| 0 | 0 | 1 | 1 | 0 | 0 | 1 | 1 |
| 0 | 0 | 1 | 1 | 0 | 0 | 1 | 1 |
| 1 | 1 | 1 | 1 | 1 | 1 | 1 | 1 |
| 1 | 1 | 1 | 1 | 1 | 1 | 1 | 1 |
| 0 | 0 | 1 | 1 | 0 | 0 | 1 | 1 |
| 0 | 0 | 1 | 1 | 0 | 0 | 1 | 1 |

→

|       | 1     | $d_1$ | $d_2$ | $d_3$ | $d_4$ | $d_5$ | $d_6$ | $d_7$ |
|-------|-------|-------|-------|-------|-------|-------|-------|-------|
| 1     | 1     | $d_1$ | $d_2$ | $d_3$ | $d_4$ | $d_5$ | $d_6$ | $d_7$ |
| $d_1$ | $d_1$ | 1     | $d_3$ | $d_2$ | $d_5$ | $d_4$ | $d_7$ | $d_6$ |
| $d_2$ | $d_2$ | $d_3$ | 0     | 0     | $d_6$ | $d_7$ | 0     | 0     |
| $d_3$ | $d_3$ | $d_2$ | 0     | 0     | $d_7$ | $d_6$ | 0     | 0     |
| $d_4$ | $d_4$ | $d_5$ | $d_6$ | $d_7$ | 1     | $d_1$ | $d_2$ | $d_3$ |
| $d_5$ | $d_5$ | $d_4$ | $d_7$ | $d_6$ | $d_1$ | 1     | $d_3$ | $d_2$ |
| $d_6$ | $d_6$ | $d_7$ | 0     | 0     | $d_2$ | $d_3$ | 0     | 0     |
| $d_7$ | $d_7$ | $d_6$ | 0     | 0     | $d_3$ | $d_2$ | 0     | 0     |

Read the third position of triplets:

| | | | | | | | | | | 1 | $d_1$ | $d_2$ | $d_3$ | $d_4$ | $d_5$ | $d_6$ | $d_7$ |
|---|---|---|---|---|---|---|---|---|---|---|---|---|---|---|---|---|---|
| C | G | C | G | C | G | C | G | | 1 | 1 | 1 | 1 | 1 | 1 | 1 | 1 | 1 | $d_1$ | $d_2$ | $d_3$ | $d_4$ | $d_5$ | $d_6$ | $d_7$ |
| T | A | T | A | T | A | T | A | | 0 | 1 | 0 | 1 | 0 | 1 | 0 | 1 | $d_1$ | $d_1$ | 0 | $d_3$ | 0 | $d_5$ | 0 | $d_7$ | 0 |
| C | G | C | G | C | G | C | G | | 1 | 1 | 1 | 1 | 1 | 1 | 1 | 1 | $d_2$ | $d_2$ | $d_3$ | 1 | $d_1$ | $d_6$ | $d_7$ | $d_4$ | $d_5$ |
| T | A | T | A | T | A | T | A | | 0 | 1 | 0 | 1 | 0 | 1 | 0 | 1 | $d_3$ | $d_3$ | 0 | $d_1$ | 0 | $d_7$ | 0 | $d_5$ | 0 |
| C | G | C | G | C | G | C | G | → | 1 | 1 | 1 | 1 | 1 | 1 | 1 | 1 | → $d_4$ | $d_4$ | $d_5$ | $d_6$ | $d_7$ | 1 | $d_1$ | $d_2$ | $d_3$ |
| T | A | T | A | T | A | T | A | | 0 | 1 | 0 | 1 | 0 | 1 | 0 | 1 | $d_5$ | $d_5$ | 0 | $d_7$ | 0 | $d_1$ | 0 | $d_3$ | 0 |
| C | G | C | G | C | G | C | G | | 1 | 1 | 1 | 1 | 1 | 1 | 1 | 1 | $d_6$ | $d_6$ | $d_7$ | $d_4$ | $d_5$ | $d_2$ | $d_3$ | 1 | $d_1$ |
| T | A | T | A | T | A | T | A | | 0 | 1 | 0 | 1 | 0 | 1 | 0 | 1 | $d_7$ | $d_7$ | 0 | $d_5$ | 0 | $d_3$ | 0 | $d_1$ | 0 |

Figure 18. The genomatrix [C G; T A]$^{(3)}$ (on the upper row) and the following four cases of its numeric representation under the denotation C=G=A=1 and T=0: 1) the second row: all positions of triplets are taken into account, and each triplet is represented as a result of multiplication of numbers 0 and 1 in accordance with its letters; 2) the third row: only the first position of triplets is taken into account; 3) the fourth row: only the second position of triplets is taken into account; 4) the fifth row: only the third position of triplets is taken into account. In each row (except for the first), the right column contains a multiplication table for the individual set of eight sparse matrices $d_0$, $d_1$, $d_2$, $d_3$, $d_4$, $d_5$, $d_6$, $d_7$ of the dyadic-shift decomposition of the relevant numeric matrix (in each row $d_0$ is the unit matrix). Each of these four multiplication tables coincides with a multiplication table of a 8-dimensional algebra which is an extension of the 2-dimensional algebra of dual numbers (Figure 16, on the right). Black (white) cells contain entries 1 (0).

## 7 Fibonacci matrices and their extensions in matrix genetics

In this section we analyze the same genomatrices but using another variant of the molecular-reasonable denotation of the nitrogenous bases: C=G=T=1 and A=0. Under this denotation, the Kronecker family of genomatrices [C G; T A]$^{(n)}$ is close connected with a family of ($2^n*2^n$)-matrices whose entries are equal to Fibonacci numbers which are multiplied by a total factor. Such matrices we will term "Fibonacci matrices" in a wide sense.

According to the author, the relationships of the genetic alphabets with dual numbers and with the Fibonacci matrices are interesting because of the following four reasons:
1) Dual numbers play a main role in screw theory which can be used to analyze spiral movements and constructions (http://en.wikipedia.org/wiki/Screw_theory);
2) Spiral formations exist in a huge number of biological structures at all levels and branches of biological organization. J.W. von Goethe called spirals as "the curves of life", and the known book [Cook, 1914] about biological spirals has also the title "The curves of life";
3) Spiral formations in living matter possess two significant differences from the spirals in inanimate nature: a) they are inherited genetically from generation to generation and for this reason they play frequently an essential role in the taxonomy of living organisms; b) they are connected with Fibonacci numbers in accordance with phyllotaxis laws (on phyllotaxis laws see for example the book [Jean, 1995]);
4) Phyllotaxis laws, which are connected with Fibonacci numbers, are one of the most known objects of mathematical biology for a long time. Hundreds of publications are devoted to this topic (for example, the book [Jean, 1995] contains more than 900 references on phyllotaxis laws).

The series of Fibonacci numbers $F_n$ (where n = 0, 1, 2, 3, …) begins with the numbers 0 and 1. Each next member of this series is equal to the sum of two previous members: $F_{n+2} = F_n + F_{n+1}$. Fibonacci numbers are used widely in the theory of optimization and in many other fields. One can find a rich collection of data about the Fibonacci numbers and the golden section on the web site of "The museum of harmony and the golden section" by A. Stakhov (www.goldenmuseum.com) and in works (Balakshin, 2008, 2011; Jean, 1995; Kappraff, 1990, 1992). Fibonacci sequences and their generalizations provide fast algorithms of discrete orthogonal transformations in numeration systems with irrational basis. Figure 19 shows the first numbers $F_n$ of the Fibonacci series.

| $n$ | 0 | 1 | 2 | 3 | 4 | 5 | 6 | 7 | 8 | 9 | 10 | 11 | 12 | 13 | 14 | ... |
|---|---|---|---|---|---|---|---|---|---|---|---|---|---|---|---|---|
| $F_n$ | 0 | 1 | 1 | 2 | 3 | 5 | 8 | 13 | 21 | 34 | 55 | 89 | 144 | 233 | 377 | ... |

Figure 19. The Fibonacci series

This series of Fibonacci numbers possesses an interesting connection with the so called Fibonacci (2*2)-matrix F= [1 1; 1 0], which are known for a long time. Exponentiation of such (2x2)-matrix in the power "n" produces new (2x2)-matrices, the components of which are equal to three adjacent Fibonacci numbers $F_{n-1}$, $F_n$, $F_{n+1}$ (Figure 20).

$$F^n = \begin{vmatrix} 1 & 1 \\ 1 & 0 \end{vmatrix}^n = \begin{vmatrix} F_{n+1} & F_n \\ F_n & F_{n-1} \end{vmatrix}$$

Figure 20. Exponentiation of the Fibonacci matrix [1 1; 1 0] produces Fibonacci numbers

Consequently in this matrix approach each set of three adjacent Fibonacci numbers $F_{n-1}$, $F_n$, $F_{n+1}$ is defined by one number "n", which is a power of a corresponding Fibonacci matrix. In view of this, one can define the Fibonacci numbers by means of these Fibonacci matrices without using the traditional algorithm of addition of two adjacent Fibonacci numbers at all. For example, the Fibonacci number $F_n$ (beginning with n = 3) can be defined as a middle component among all components of the matrix $F^n$. Or the Fibonacci number $F_n$ can be defined as that number which is a component of each of the three matrices $F^{n-1}$, $F^n$ and $F^{n+1}$ simultaneously.

Eigenvalues of the Fibonacci matrix [1 1; 1 0] are equal to the golden section "1.618…" and its reverse magnitude "-0.618…" with the sign "-". This feature allows one to propose new definition of the golden section: the golden section is the positive eigenvalue of the Fibonacci matrix [Petoukhov, 2008a; Petoukhov, He, 2010]. Kronecker exponentiation of the Fibonacci matrix $[1\ 1;\ 1\ 0]^{(n)}$ (or $[0\ 1;\ 1\ 1]^{(n)}$), where n = 2, 3, 4,..., gives the $(2^n*2^n)$-matrices, all eigenvalues of which are equal to the golden section in different integer powers only (magnitudes of the eigenvalues can be positive or negative, of course).

Generalizations of the Fibonacci (2*2)-matrices in a form of families of $(2^n×2^n)$-matrices are possible (Petoukhov, 2003-2004, 2008; Petoukhov, He, 2010, Chapter 10). Let us return to the (4*4)-genomatrix [C G; T A]$^{(2)}$ of the 16 duplets (Figure 17).

Figure 21 shows Fibonacci (4*4)-matrices $V_1$ and $V_2$ which are received from the (4*4)-genomatrix [C G; T A]$^{(2)}$ which is represented on this Figure in forms $W_1$ and $W_2$ where only one of the two positions (the first or the second) of each duplet is taken into account. Using the molecular designation C=G=T=1 and A=0 leads to the two numeric presentations $V_1$ and $V_2$ of the genomatrices $W_1$ and $W_2$, each of which is a Fibonacci (4*4)-matrix because its exponentiation gives the matrix $V_1^n$ or the matrix $V_2^n$, all components of which are three adjacent Fibonacci numbers $F_{n-1}$, $F_n$, $F_{n+1}$ with a factor $2^{n-1}$.

$$[C\ G;\ T\ A]^{(2)} = \begin{array}{|c|c|c|c|} \hline CC & CG & GC & GG \\ \hline CT & CA & GT & GA \\ \hline TC & TG & AC & AG \\ \hline TT & TA & AT & AA \\ \hline \end{array} \rightarrow W_1 = \begin{array}{|c|c|c|c|} \hline C & C & G & G \\ \hline C & C & G & G \\ \hline T & T & A & A \\ \hline T & T & A & A \\ \hline \end{array} \rightarrow V_1 = \begin{array}{|c|c|c|c|} \hline 1 & 1 & 1 & 1 \\ \hline 1 & 1 & 1 & 1 \\ \hline 1 & 1 & 0 & 0 \\ \hline 1 & 1 & 0 & 0 \\ \hline \end{array}$$

$$W_2 = \begin{array}{|c|c|c|c|} \hline C & G & C & G \\ \hline T & A & T & A \\ \hline C & G & C & G \\ \hline T & A & T & A \\ \hline \end{array} \rightarrow V_2 = \begin{array}{|c|c|c|c|} \hline 1 & 1 & 1 & 1 \\ \hline 1 & 0 & 1 & 0 \\ \hline 1 & 1 & 1 & 1 \\ \hline 1 & 0 & 1 & 0 \\ \hline \end{array}$$

$$V_1^n = 2^{n-1} * \begin{array}{|c|c|c|c|} \hline F_{n+1} & F_{n+1} & F_n & F_n \\ \hline F_{n+1} & F_{n+1} & F_n & F_n \\ \hline F_n & F_n & F_{n-1} & F_{n-1} \\ \hline F_n & F_n & F_{n-1} & F_{n-1} \\ \hline \end{array} \quad ; \quad V_2^n = 2^{n-1} * \begin{array}{|c|c|c|c|} \hline F_{n+1} & F_n & F_{n+1} & F_n \\ \hline F_n & F_{n-1} & F_n & F_{n-1} \\ \hline F_{n+1} & F_n & F_{n+1} & F_n \\ \hline F_n & F_{n-1} & F_n & F_{n-1} \\ \hline \end{array}$$

Figure 21. On the two upper rows: the genomatrix [C A; G T]$^{(2)}$ with 16 duplets and the genomatrices $W_1$ ($W_2$) where only the first (the second) position of each duplet is taken into account. The representations $V_1$ and $V_2$ of the genomatrices $W_1$ and $W_2$ under the molecular designation A=C=G=1 and T=0 are also shown. Black (white) cells contain entries 1 (0). On the lower row: exponentiation of the matrices $V_1$ and $V_2$ leads to matrices with Fibonacci numbers $F_{n-1}$, $F_n$, $F_{n+1}$ and the factor $2^{n-1}$.

By analogy Figure 22 shows Fibonacci (8*8)-matrices $G_1$, $G_2$ and $G_3$ which are received from the (8*8)-genomatrix [C G; T A]$^{(3)}$ if only one of the three positions (the first, the second or the third position) of each triplet is taken into account. Using the molecular designation C=G=T=1 and A=0 leads to the three numeric representations $G_1$, $G_2$ and $G_3$ each of which is a Fibonacci (8*8)-matrix because its exponentiation gives the matrix $G_1^n$, $G_2^n$ and $G_3^n$, all components of which are three adjacent Fibonacci numbers $F_{n-1}$, $F_n$, $F_{n+1}$ with a factor $4^{n-1}$.

| C | C | C | C | G | G | G | G |
|---|---|---|---|---|---|---|---|
| C | C | C | C | G | G | G | G |
| C | C | C | C | G | G | G | G |
| C | C | C | C | G | G | G | G |
| T | T | T | T | A | A | A | A |
| T | T | T | T | A | A | A | A |
| T | T | T | T | A | A | A | A |
| T | T | T | T | A | A | A | A |

;

| C | C | G | G | C | C | G | G |
|---|---|---|---|---|---|---|---|
| C | C | G | G | C | C | G | G |
| T | T | A | A | T | T | A | A |
| T | T | A | A | T | T | A | A |
| C | C | G | G | C | C | G | G |
| C | C | G | G | C | C | G | G |
| T | T | A | A | T | T | A | A |
| T | T | A | A | T | T | A | A |

;

| C | G | C | G | C | G | C | G |
|---|---|---|---|---|---|---|---|
| T | A | T | A | T | A | T | A |
| C | G | C | G | C | G | C | G |
| T | A | T | A | T | A | T | A |
| C | G | C | G | C | G | C | G |
| T | A | T | A | T | A | T | A |
| C | G | C | G | C | G | C | G |
| T | A | T | A | T | A | T | A |

$G_1 = \begin{array}{|cccccccc|} \hline 1 & 1 & 1 & 1 & 1 & 1 & 1 & 1 \\ 1 & 1 & 1 & 1 & 1 & 1 & 1 & 1 \\ 1 & 1 & 1 & 1 & 1 & 1 & 1 & 1 \\ 1 & 1 & 1 & 1 & 1 & 1 & 1 & 1 \\ 1 & 1 & 1 & 1 & 0 & 0 & 0 & 0 \\ 1 & 1 & 1 & 1 & 0 & 0 & 0 & 0 \\ 1 & 1 & 1 & 1 & 0 & 0 & 0 & 0 \\ 1 & 1 & 1 & 1 & 0 & 0 & 0 & 0 \\ \hline \end{array}$ ; $G_2 = \begin{array}{|cccccccc|} \hline 1 & 1 & 1 & 1 & 1 & 1 & 1 & 1 \\ 1 & 1 & 1 & 1 & 1 & 1 & 1 & 1 \\ 1 & 1 & 0 & 0 & 1 & 1 & 0 & 0 \\ 1 & 1 & 0 & 0 & 1 & 1 & 0 & 0 \\ 1 & 1 & 1 & 1 & 1 & 1 & 1 & 1 \\ 1 & 1 & 1 & 1 & 1 & 1 & 1 & 1 \\ 1 & 1 & 0 & 0 & 1 & 1 & 0 & 0 \\ 1 & 1 & 0 & 0 & 1 & 1 & 0 & 0 \\ \hline \end{array}$ ; $G_3 = \begin{array}{|cccccccc|} \hline 1 & 1 & 1 & 1 & 1 & 1 & 1 & 1 \\ 1 & 0 & 1 & 0 & 1 & 0 & 1 & 0 \\ 1 & 1 & 1 & 1 & 1 & 1 & 1 & 1 \\ 1 & 0 & 1 & 0 & 1 & 0 & 1 & 0 \\ 1 & 1 & 1 & 1 & 1 & 1 & 1 & 1 \\ 1 & 0 & 1 & 0 & 1 & 0 & 1 & 0 \\ 1 & 1 & 1 & 1 & 1 & 1 & 1 & 1 \\ 1 & 0 & 1 & 0 & 1 & 0 & 1 & 0 \\ \hline \end{array}$

$$G_1^n = 4^{n-1} * \begin{vmatrix} F_{n+1} & F_{n+1} & F_{n+1} & F_{n+1} & F_n & F_n & F_n & F_n \\ F_{n+1} & F_{n+1} & F_{n+1} & F_{n+1} & F_n & F_n & F_n & F_n \\ F_{n+1} & F_{n+1} & F_{n+1} & F_{n+1} & F_n & F_n & F_n & F_n \\ F_{n+1} & F_{n+1} & F_{n+1} & F_{n+1} & F_n & F_n & F_n & F_n \\ F_n & F_n & F_n & F_n & F_{n-1} & F_{n-1} & F_{n-1} & F_{n-1} \\ F_n & F_n & F_n & F_n & F_{n-1} & F_{n-1} & F_{n-1} & F_{n-1} \\ F_n & F_n & F_n & F_n & F_{n-1} & F_{n-1} & F_{n-1} & F_{n-1} \\ F_n & F_n & F_n & F_n & F_{n-1} & F_{n-1} & F_{n-1} & F_{n-1} \end{vmatrix} \; ; \; G_2^n = 4^{n-1} * \begin{vmatrix} F_{n+1} & F_{n+1} & F_n & F_n & F_{n+1} & F_{n+1} & F_n & F_n \\ F_{n+1} & F_{n+1} & F_n & F_n & F_{n+1} & F_{n+1} & F_n & F_n \\ F_n & F_n & F_{n-1} & F_{n-1} & F_n & F_n & F_{n-1} & F_{n-1} \\ F_n & F_n & F_{n-1} & F_{n-1} & F_n & F_n & F_{n-1} & F_{n-1} \\ F_{n+1} & F_{n+1} & F_n & F_n & F_{n+1} & F_{n+1} & F_n & F_n \\ F_{n+1} & F_{n+1} & F_n & F_n & F_{n+1} & F_{n+1} & F_n & F_n \\ F_n & F_n & F_{n-1} & F_{n-1} & F_n & F_n & F_{n-1} & F_{n-1} \\ F_n & F_n & F_{n-1} & F_{n-1} & F_n & F_n & F_{n-1} & F_{n-1} \end{vmatrix}$$

$$G_3^n = 4^{n-1} * \begin{vmatrix} F_{n+1} & F_n & F_{n+1} & F_n & F_{n+1} & F_n & F_{n+1} & F_n \\ F_n & F_{n-1} & F_n & F_{n-1} & F_n & F_{n-1} & F_n & F_{n-1} \\ F_{n+1} & F_n & F_{n+1} & F_n & F_{n+1} & F_n & F_{n+1} & F_n \\ F_n & F_{n-1} & F_n & F_{n-1} & F_n & F_{n-1} & F_n & F_{n-1} \\ F_{n+1} & F_n & F_{n+1} & F_n & F_{n+1} & F_n & F_{n+1} & F_n \\ F_n & F_{n-1} & F_n & F_{n-1} & F_n & F_{n-1} & F_n & F_{n-1} \\ F_{n+1} & F_n & F_{n+1} & F_n & F_{n+1} & F_n & F_{n+1} & F_n \\ F_n & F_{n-1} & F_n & F_{n-1} & F_n & F_{n-1} & F_n & F_{n-1} \end{vmatrix}$$

Figure 22 On the upper row: transformations of the genomatrix [C G; T A]$^{(3)}$ of 64 triplets (Figure 18) into three genomatrices each of which contains only letters on a certain (the first, the second or the third) position of each triplet. On the second row: representations $G_1$, $G_2$ and $G_3$ of these three genomatrices in numeric forms under the denotation C=G=T=1 and A=0. Black (white) cells contain entries 1 (0). On the lower rows: all entries of the Fibonacci matrices $G_1^n$, $G_2^n$ and $G_3^n$ are three adjacent Fibonacci numbers $F_{n-1}$, $F_n$, $F_{n+1}$ with a total factor $4^{n-1}$.

The following fact, which was revealed by the author, is interesting for modelling phyllotaxis laws by means of the mathematical interrelations of Fibonacci matrices with dual numbers and screw theory. Usual multiplications of the Fibonacci matrix F = [1 1; 1 0] with the matrix presentation of the dual number $D_1$ = [1 1; 0 1] (or $D_2$ = [1 0; 1 1]) gives new types of Fibonacci (2*2)-matrices whose exponentiations lead to matrices with Fibonacci numbers $F_n$ inside them (Figure 23).

$$D_1 * F^n = \begin{vmatrix} 1 & 0 \\ 1 & 1 \end{vmatrix} * \begin{vmatrix} 1 & 1 \\ 1 & 0 \end{vmatrix}^n = \begin{vmatrix} F_{n+1} & F_n \\ F_{n+2} & F_{n+1} \end{vmatrix} ; \quad F^n * D_1 = \begin{vmatrix} 1 & 1 \\ 1 & 0 \end{vmatrix}^n * \begin{vmatrix} 1 & 0 \\ 1 & 1 \end{vmatrix} = \begin{vmatrix} F_{n+2} & F_n \\ F_{n+1} & F_{n-1} \end{vmatrix}$$

$$D_1 * F^n * D_1 = \begin{vmatrix} 1 & 0 \\ 1 & 1 \end{vmatrix} * \begin{vmatrix} 1 & 1 \\ 1 & 0 \end{vmatrix}^n * \begin{vmatrix} 1 & 0 \\ 1 & 1 \end{vmatrix} = \begin{vmatrix} F_{n+2} & F_n \\ F_{n+3} & F_{n+1} \end{vmatrix}$$

Figure 23. Combinations of the classic Fibonacci (2*2)-matrix F=[1 1; 1 0]$^{(n)}$ with the matrix representation of the dual number $D_1$=[1 0; 1 1] give new types of Fibonacci matrices whose entries are equal to Fibonacci numbers

Pairs of eigenvalues of the matrices $D_1*F$, $F*D_1$ and $D_1*F*D_1$ (Figure 24) are interconnected in each case by means of inversion of their absolute values: $2.4142 = (0.4142)^{-1}$, $3.3028 = (0.3028)^{-1}$. It seems that such type of interconnections holds true also for pairs of eigenvalues of the matrices $D_1*F^n$, $F^n*D_1$ and $D_1*F^n*D_1$.

| Matrix | Eigenvalues |
|---|---|
| $D_1*F$ | 2.4142, -0.4142 |
| $F*D_1$ | -0.4142, 2.4142 |
| $D_1*F*D_1$ | 3.3028, -0.3028 |

Figure 24. Eigenvalues of the matrices $D_1*F$, $F^n*D_1$ and $D_1*F^n*D_1$ from Figure 23

But what one can say on Kronecker multiplications of the Fibonacci matrix F=[1 1; 1 0] with the matrix representation of the dual number $D_1$ = [1 0; 1 1] (or $D_2$ = [1 1; 0 1])? It is interesting that in this case we receive new Fibonacci's types of (4*4)-matrices with Fibonacci numbers $F_n$ inside them (Figure 25).

$$(D_1 \otimes F^n)^k = \begin{array}{|c|c|c|c|} \hline F_{k*n+1} & F_{k*n} & 0 & 0 \\ \hline F_{k*n} & F_{k*n-1} & 0 & 0 \\ \hline k*F_{k*n+1} & k*F_{k*n} & F_{k*n+1} & F_{k*n} \\ \hline k*F_{k*n} & k*F_{k*n-1} & F_{k*n} & F_{k*n-1} \\ \hline \end{array}$$

$$(F^n \otimes D_1)^k = \begin{array}{|c|c|c|c|} \hline F_{k*n+1} & 0 & F_{k*n} & 0 \\ \hline k*F_{k*n+1} & F_{k*n+1} & k*F_{k*n} & F_{k*n} \\ \hline F_{k*n} & 0 & F_{k*n-1} & 0 \\ \hline k*F_{k*n} & F_{k*n} & k*F_{k*n-1} & F_{k*n-1} \\ \hline \end{array}$$

Figure 25. New Fibonacci's types of (4*4)-matrices on the base of Kronecker multiplication of the classic Fibonacci matrices F=[1 1; 1 0]$^{(n)}$ and the matrix representation $D_1$ of the dual number [1 1; 0 1]. The symbol $\otimes$ means Kronecker multiplication. $F_n$ means Fibonacci numbers from Figure 19. The entry "0" is equal to $F_0$.

## 8. Genetic matrices and bicomplex numbers

In this section we will consider the same genetic matrices of the Kronecker family [C G; T A], [C G; T A]$^{(2)}$ and [C G; T A]$^{(3)}$ (Figures 17 and 18) but under another molecular denotation A=C=G=1 and T=-1 to receive their numeric representations by means of the following algorithm: each duplet and each triplet is interpreted as product of numbers "+1" and "-1" taking into account the mentioned denotation. For example the duplet TC is replaced by number (-1)*(+1) = -1 and the triplet GTT is replaced by number (+1)*(-1)*(-1) = +1. Figure 26 shows that under this denotation the genomatrix [C G; T A] coinsides with a matrix representation K of the complex number x+i*y whose coordinates are equal to 1 (x=y=1, $i^2$ = -1). The dyadic-shift decomposition of the matrix K gives a set of two sparse matrices, which is closed in relation to multiplication. It defines a multiplication table, which coincides with the multiplication table of 2-dimensional algebra of complex numbers (Figure 26, on the right).

| C | G |
|---|---|
| T | A |

$\rightarrow$ K=1+i = $\begin{array}{|c|c|} \hline 1 & 1 \\ \hline -1 & 1 \\ \hline \end{array}$ = $\begin{array}{|c|c|} \hline 1 & 0 \\ \hline 0 & 1 \\ \hline \end{array}$ + $\begin{array}{|c|c|} \hline 0 & 1 \\ \hline -1 & 0 \\ \hline \end{array}$ ;

| | 1 | i |
|---|---|---|
| 1 | 1 | i |
| i | i | -1 |

Figure 26 The numeric representation of the genetic matrix [C G; T A] in the case of the denotation A=C=G=1 and T=-1. The matrix K is the matrix representation of the complex number 1+i (where $i^2$=-1). Black (white) cells contain entries +1 (-1). The multiplication table of complex numbers is shown on the right.

Figure 27 shows the genomatrix [C G; T A]$^{(2)}$ and its numeric representation B under the denotation C=G=A=1 and T=-1 in the three cases: 1) both positions of duplets are taken into account, and each duplet is represented as a result of multiplication of numbers "+1" and "-1" in accordance with its letters; the numeric (4*4)-matrix in this case is a Hadamard matrix additionally; 2) only the first position of duplets is taken into account; 3) only the second position of duplets is taken into account. The dyadic-shift decomposition of each numeric representation of matrices on Figure 27 gives a set of four sparse matrices $b_0$, $b_1$, $b_2$, $b_3$ (this set is individual for each of the three cases) which is closed in relation to multiplication; its multiplication table corresponds to a 4-dimensional algebra of bicomplex numbers. Bicomplex numbers are an extension of complex numbers and they are well known in mathematics and its applications in many fields (see for example [Babadag, 2009; Rochon, 2000; Rochon, Shapiro, 2003; Ronn, 2001; Smirnov, 2005; Tremblay, 2005]). In this reason the genomatrices B, $B_1$ and $B_2$ on Figure 27 are matrix presentations of the bicomplex number $a+b*b_1+c*b_2+d*b_3$ whose coordinates are equal to 1 (a=b=c=d=1).

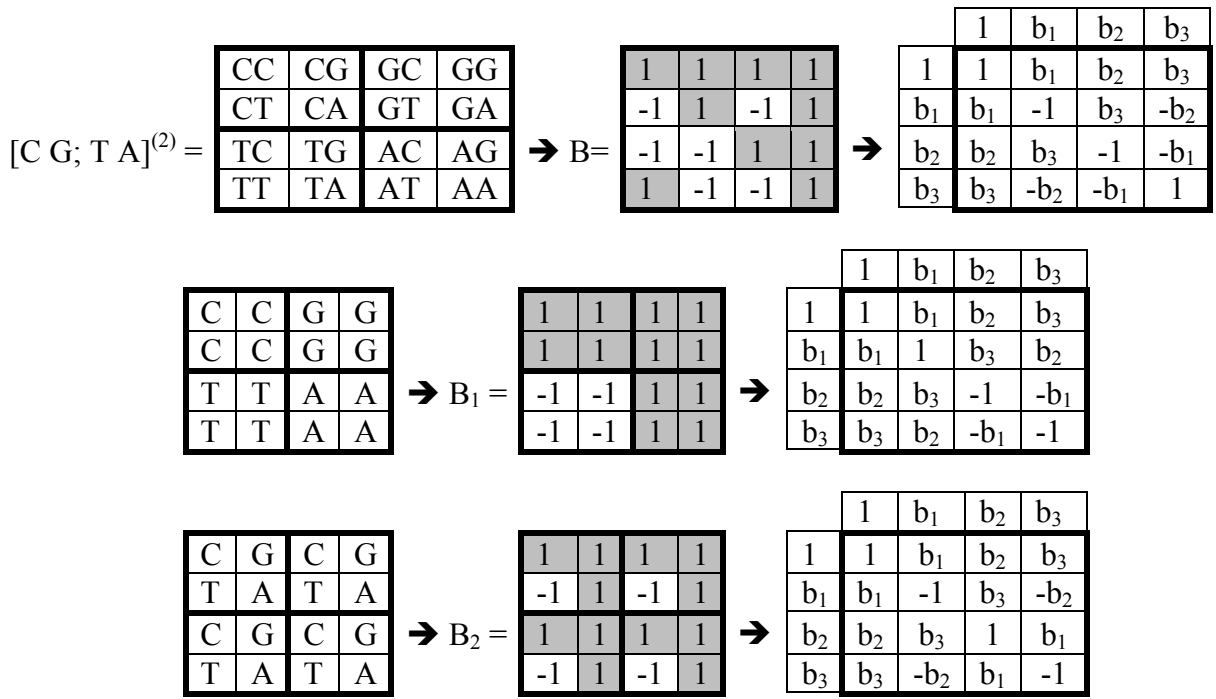

Figure 27. The genomatrix $[C\ G;\ T\ A]^{(2)}$ and the following three cases of its numeric representation under the denotation C=G=A=1 and T=-1: 1) the upper row: both positions of duplets are taken into account, and each duplet is represented as a result of multiplication of numbers "+1" and "-1" in accordance with its letters; the numeric matrix B is the Hadamard matrix additionally; 2) the middle row: only the first position of duplets is taken into account; 3) the lower row: only the second position of duplets is taken into account. In each row, the right column contains a multiplication table for the appropriate set of four sparse matrices $b_0$, $b_1$, $b_2$, $b_3$ of the dyadic-shift decomposition of the relevant numeric matrix (in each row $b_0$ is the unit matrix). Each of these three multiplication tables is an extension of the multiplication table of the 2-dimensional algebra of complex numbers (Figure 26, on the right). Black (white) cells contain entries 1 (-1).

Figure 28 shows the genomatrix $[C\ G;\ T\ A]^{(3)}$ of 64 triplets and its numeric representation under the same denotation C=G=A=1 and T= -1 in the four cases: 1) all positions of triplets are taken into account, and each triplet is represented as a result of multiplication of numbers "+1" and "-1" in accordance with its letters (for example the triplet CTA is represented by "-1" because of 1*(-1)*1=-1); the numeric (8*8)-matrix in this case is a Hadamard matrix additionally;  2) only the first position of triplets is taken into account; 3) only the second position of triplets is taken into account; 4) only the third position of triplets is taken into account. The dyadic-shift decomposition of each numeric representation of matrices on Figure 28 gives a set of eight sparse matrices $b_0$, $b_1$, $b_2$, $b_3$, $b_4$, $b_5$, $b_6$, $b_7$ (this set is individual for each of the four cases but $b_0$ is the unit matrix in all sets) which is closed in relation to multiplication; its multiplication table corresponds to a case of 8-dimensional algebra which is an extension of the 2-dimensional algebra of complex numbers.

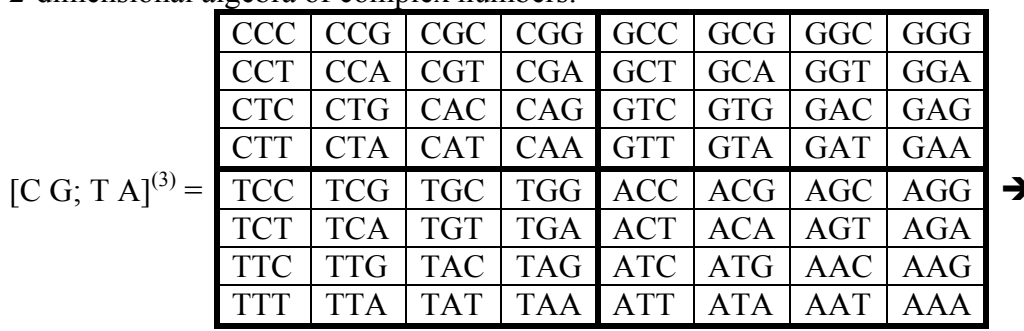

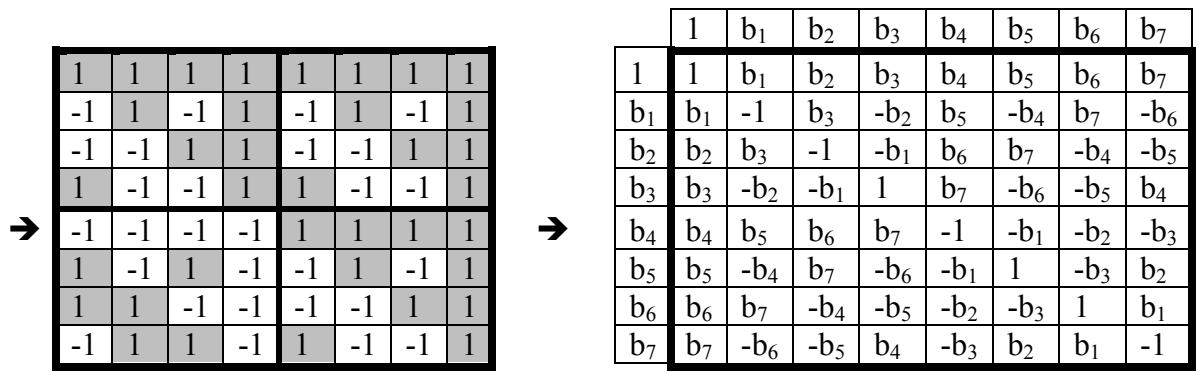

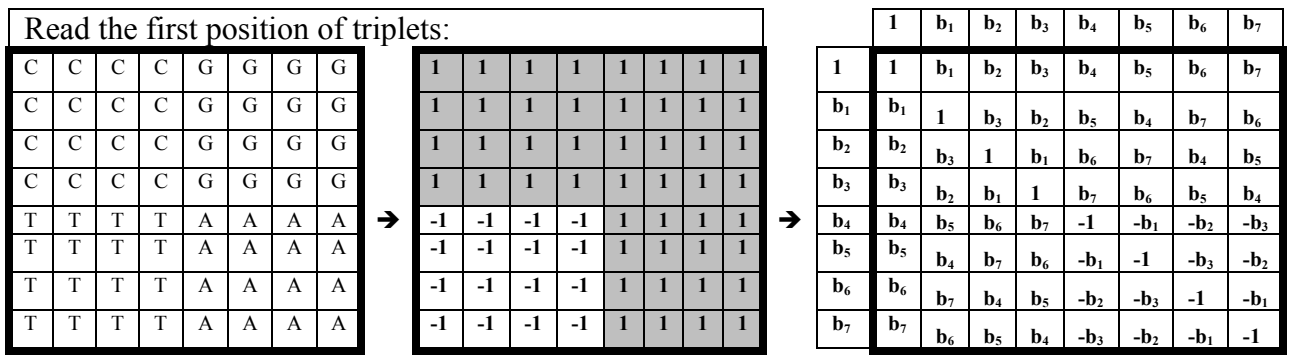

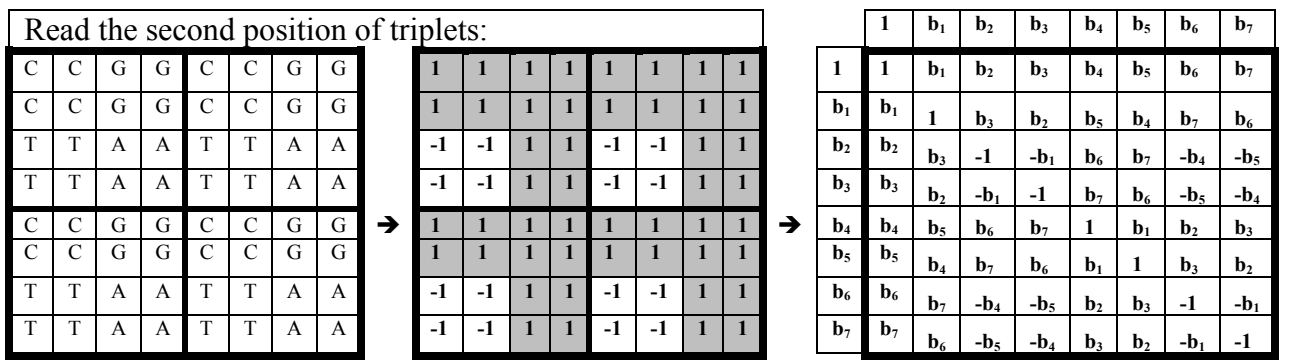

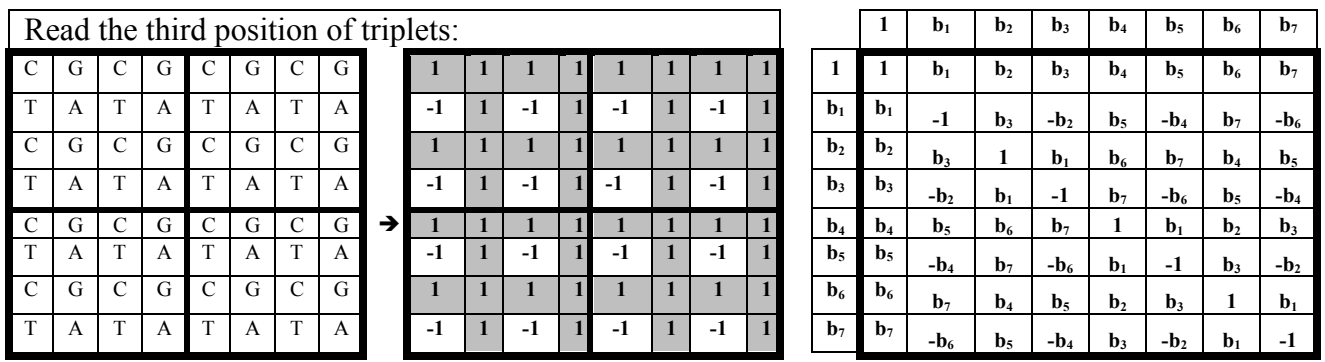

Figure 28. The genomatrix [C G; T A]$^{(3)}$ (on the upper row) and the following four cases of its numeric representation under the denotation C=G=A=1 and T= -1: 1) the second row: all positions of triplets are taken into account, and each triplet is represented as a result of multiplication of numbers "+1" and "-1" in accordance with its letters (the numeric matrix in this row is a Hadamard matrix additionally); 2) the third row: only the first position of triplets is taken into account; 3) the fourth row: only the second position of triplets is taken into account; 4) the fifth row: only the third position of triplets is taken into account. In each row (except for the first), the right column contains a multiplication table for the individual set of eight sparse matrices $b_0$, $b_1$, $b_2$, $b_3$, $b_4$, $b_5$, $b_6$, $b_7$ of the dyadic-shift decomposition of the relevant numeric matrix (in each row $b_0$ is the unit matrix). Each of these four multiplication tables coincides with a multiplication table of an 8-dimensional algebra, which is an extension of the 2-dimensional algebra of dual numbers (Figure 16, on the right). Black (white) cells contain entries 1 (0).

The multiplication table on the second row of Figure 28 corresponds to the multiplication table of the 8-dimensional algebra of bicomplex numbers over field of complex numbers (this extension of bicomplex numbers can be termed bi-bicomplex numbers). In this reason the numeric (8*8)-genomatrix in the second row is the bi-bicomplex number whose coordinates are equal to 1. Figure 29 shows the interconnection between the numeric (4*4)-genomatrix B (Figure 27) and the (8*8)-genomatrix $T_{CAGT}$ (Figure 27) on the base of Kronecker multiplication of B with the matrix presentation [1 -1; 1 1] of the complex number x+i*y whose coordinates are equal to 1 (x=y=1).

Figure 29. The interconnection between the (4*4)-genomatrix B (Figure 27) and the numeric (8*8)-genomatrix (Figure 28, on the second row) on the base of Kronecker multiplication of the matrix B with the matrix presentation [1 1; -1 1] of the complex number whose coordinates are equal to 1. Here all the three matrices are Hadamard matrices

The same approach can be applied to analyze interconnections of hypercomplex numbers with other genetic (2*2)-matrices in their Kronecker powers.

### 9. Hypercomplex numbers as informational essence

Hypercomplex numbers are widely used in geometry and physics. Therefore, they usually are thought as the tools of geometry and physics. The author would like to emphasize the fact that a specifity of matrix representations of hypercomplex numbers (such as bicomplex numbers, Hamilton quaternions, split-quaternions by Cockle and their 8-dimensional extensions which were considered in our previous article [Petoukhov, 2012]) testifies, that these hypercomplex numbers are deeply connected with informatics or, in other words, that they are information essense in their basis. This specifity is expressed in the two following facts:
- All the coordinates in matrix representations of these hypercomplex numbers are disposed in accordance with structure of diadic-shift matrices (Figure 1) that is in accordance with the mathematical operation of modulo-2 addition, which is one of the main operations in computer informatics. Figure 30 shows identical dyadic-shift dispositions of coordinates a, b, c, d in matrix representations of different 4-dimensional hypercomplex numbers: each type of coordinates is disposed inside cells with an individual diadic-shift numeration (compare with Figure 1). These representations differ from each other only by dispositions of signs "+" and "-" of different coordinates, placed inside matrices identically. This dyadic-shift principle of matrix composition holds true also for different matrix representations of 8-dimensional extensions of these 4-dimensional numbers;
- If all the coordinates of such hypercomplex numbers are equal to 1 (a=b=c=…=1), each of these matrix representations on Figure 30 becomes a Hadamard matrix $H_n$ which satisfies the condition $H_n * H_n^T = n*E$ where $H_n^T$ is a transposed matrix, E – the unity matrix. In other words, these hypercomplex numbers have a hidden connection with Hadamard matrices and with their orthogonal systems of Walsh functions, which are well-known effective tools in information technology and noise-immune coding.

The author doesn't know literature sources where such statement about information essence of these hypercomplex numbers was published. Perhaps it is a new one.

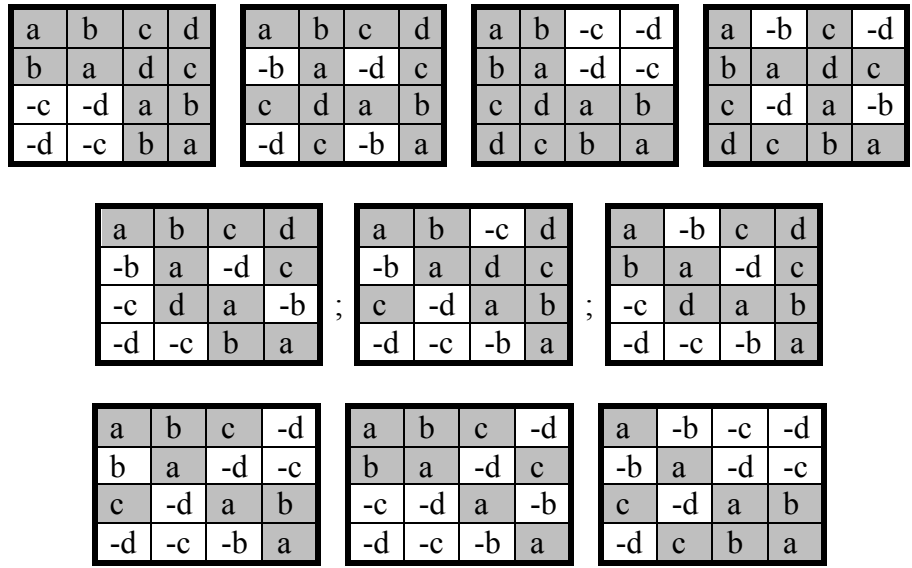

Figure 30. Some examples of different matrix representations of the following
4-dimensional hypercomplex numbers: 1) on the upper row: bicomplex numbers;
2) on the middle row: Hamilton quaternions; 3) on the lower row: split-quaternions
by Cockle [Petoukhov, 2012]

## 10. Genetic matrices with internal complementarities

This section is devoted to the genetic matrices [C T; A G]$^{(2)}$ and [C T; A G]$^{(3)}$ and their natural numeric presentations in forms of Rademacher and Hadamard matrices. These presentations were used in our previous publications [Petoukhov, 2008a,b, 2011a,b, 2012] but this section describes new algebraic properties of these genetic matrices. In accordance with phenomenologic properties of the genetic code, the set of 16 genetic duplets is devided in two equal subsets of 8 "strong" duplets (CC, CT, CG, TC, AC, GC, GT, GG) and 8 "weak" duplets (CA, TA, TG, TT, AT, AA, AG, GA). The set of 64 triplets is devided in two equal subsets of 32 triplets with strong roots (that is with strong duplets on their two first positions) and 32 triplets with weak roots (see details in [Petoukhov, 2008a,b, 2011a,b, 2012]). Figure 31 shows genomatrices [C T; A G]$^{(2)}$ and [C T; A G]$^{(3)}$ with black-and-white mosaics where black color corresponds to 8 strong duplets and 32 triplets with strong roots.

Mosaics of each column in matrices on Figure 31 have meander-like character and correspond to Rademacher functions, which are well known in discrete signals processing; square waves of Rademacher functions have only two values of their amplitudes: +1 and -1. Correspondingly one can represent of symbolic genomatrices [C T; A G]$^{(2)}$ and [C T; A G]$^{(3)}$ in their numeric Rademacher forms $R_4$ and $R_8$ (Figure 32) where each of black (white) cells contains +1 (-1).

$[C\ T;\ A\ G]^{(2)} = $
| CC | CT | TC | TT |
|----|----|----|----|
| CA | CG | TA | TG |
| AC | AT | GC | GT |
| AA | AG | GA | GG |

; $[C\ T;\ A\ G]^{(3)} = $
| CCC | CCT | CTC | CTT | TCC | TCT | TTC | TTT |
|-----|-----|-----|-----|-----|-----|-----|-----|
| CCA | CCG | CTA | CTG | TCA | TCG | TTA | TTG |
| CAC | CAT | CGC | CGT | TAC | TAT | TGC | TGT |
| CAA | CAG | CGA | CGG | TAA | TAG | TGA | TGG |
| ACC | ACT | ATC | ATT | GCC | GCT | GTC | GTT |
| ACA | ACG | ATA | ATG | GCA | GCG | GTA | GTG |
| AAC | AAT | AGC | AGT | GAC | GAT | GGC | GGT |
| AAA | AAG | AGA | AGG | GAA | GAG | GGA | GGG |

Figure 31. Matrices $[C\ T;\ A\ G]^{(2)}$ and $[C\ T;\ A\ G]^{(3)}$ with black-and-white mosaics where black color corresponds to 8 strong duplets and 32 triplets with strong roots.

$R_4 = $
| 1  | 1  | 1  | -1 |
|----|----|----|----|
| -1 | 1  | -1 | -1 |
| 1  | -1 | 1  | 1  |
| -1 | -1 | -1 | 1  |

; $R_8 = $
| 1  | 1  | 1  | 1  | 1  | 1  | -1 | -1 |
|----|----|----|----|----|----|----|----|
| 1  | 1  | 1  | 1  | 1  | 1  | -1 | -1 |
| -1 | -1 | 1  | 1  | -1 | -1 | -1 | -1 |
| -1 | -1 | 1  | 1  | -1 | -1 | -1 | -1 |
| 1  | 1  | -1 | -1 | 1  | 1  | 1  | 1  |
| 1  | 1  | -1 | -1 | 1  | 1  | 1  | 1  |
| -1 | -1 | -1 | -1 | -1 | -1 | 1  | 1  |
| -1 | -1 | -1 | -1 | -1 | -1 | 1  | 1  |

Figure 32. Rademacher representations $R_4$ and $R_8$ of genomatrices $[C\ T;\ A\ G]^{(2)}$ and $[C\ T;\ A\ G]^{(3)}$ from Figure 31.

Let us initially analyze the matrix $R_4$. It is a sum of two matrices $R_{4L}$ and $R_{4R}$ (Figure 33).

$R_4 = R_{4L} + R_{4R} = $
| 1  | 0 | 1  | 0 |
|----|---|----|---|
| -1 | 0 | -1 | 0 |
| 1  | 0 | 1  | 0 |
| -1 | 0 | -1 | 0 |
+
| 0 | 1  | 0 | -1 |
|---|----|---|----|
| 0 | 1  | 0 | -1 |
| 0 | -1 | 0 | 1  |
| 0 | -1 | 0 | 1  |
, where

$R_{4L} = R0_{4L} + R1_{4L} = $
| 1  | 0 | 0  | 0 |
|----|---|----|---|
| -1 | 0 | 0  | 0 |
| 0  | 0 | 1  | 0 |
| 0  | 0 | -1 | 0 |
+
| 0  | 0 | 1  | 0 |
|----|---|----|---|
| 0  | 0 | -1 | 0 |
| 1  | 0 | 0  | 0 |
| -1 | 0 | 0  | 0 |
,

$R_{4R} = R0_{4R} + R1_{4R} = $
| 0 | 1 | 0 | 0 |
|---|---|---|---|
| 0 | 1 | 0 | 0 |
| 0 | 0 | 0 | 1 |
| 0 | 0 | 0 | 1 |
+
| 0 | 0  | 0 | -1 |
|---|----|---|----|
| 0 | 0  | 0 | -1 |
| 0 | -1 | 0 | 0  |
| 0 | -1 | 0 | 0  |

Figure 33. Upper row: the representation of the matrix $R_4$ as sum of matrices $R_{4L}$ and $R_{4R}$. Other rows: representations of matrices $R_{4L}$ and $R_{4R}$ as sums of matrices $R0_{4L}$, $R1_{4L}$, $R0_{4R}$ and $R1_{4R}$.

It is unexpected but the set of two (4*4)-matrices $R0_{4L}$ and $R1_{4L}$ is closed in relation to multiplication and it defines the multiplication table of these matrices (Figure 34), which is identical to the well-known multiplication table of split-complex numbers (their synonyms are

Lorentz numbers, hyperbolic numbers, perplex numbers, double numbers, etc. - http://en.wikipedia.org/wiki/Split-complex_number). Split-complex numbers are a two-dimensional commutative algebra over the real numbers.

|   | $R0_{4L}$ | $R1_{4L}$ |
|---|---|---|
| $R0_{4L}$ | $R0_{4L}$ | $R1_{4L}$ |
| $R1_{4L}$ | $R1_{4L}$ | $R0_{4L}$ |

; $D_L = a_0 * R0_{4L} + a_2 * R1_{4L} = $

| $a_0$ | 0 | $a_2$ | 0 |
|---|---|---|---|
| $-a_0$ | 0 | $-a_2$ | 0 |
| $a_2$ | 0 | $a_0$ | 0 |
| $-a_2$ | 0 | $-a_0$ | 0 |

; $D_L^{-1} = (a_0^2 - a_2^2)^{-1} *$

| $a_0$ | 0 | $-a_2$ | 0 |
|---|---|---|---|
| $-a_0$ | 0 | $a_2$ | 0 |
| $-a_2$ | 0 | $a_0$ | 0 |
| $a_2$ | 0 | $-a_0$ | 0 |

Figure 34. The multiplication table of two (4*4)-matrices $R0_{4L}$ and $R1_{4L}$ (Figure 33), which is a set of basic elements of split-complex numbers $D_L = a_0*R0_{4L}+a_2*R1_{4L}$, where $a_0$, $a_2$ are real numbers. The matrix $D_L^{-1}$ is the inverse matrix for $D_L$ if $a_0 \neq a_2$.

The set of (4*4)-matrices $D_L = a_0*R0_{4L}+a_2*R1_{4L}$, where $a_0$, $a_2$ are real numbers, represents split-complex numbers in the special (4*4)-matrix form (Figure 34). The classical identity matrix E=[1 0 0 0; 0 1 0 0; 0 0 1 0; 0 0 0 1] is absent in the set of matrices $D_L$, where the matrix $R0_{4L}$ plays a role of the real unit (identity matrix for this set). In the case $a_0 \neq a_2$, the matrix $D_L^{-1}$ (Figure 34) is the inverse matrix for $D_L$ since $D_L*D_L^{-1} = D_L^{-1}*D_L = R0_{4L}$. From this point of view, the genetic matrix $R_{4L}$ is split-complex number with unit coordinates.

One should note that the (2*2)-matrix [$a_0$ $a_2$; $a_2$ $a_0$] is usually used for a matrix representation of split-complex numbers (http://en.wikipedia.org/wiki/Split-complex_number). In the case of genetic matrices, we reveal that 4-dimensional spaces can contain 2-parametric subspaces, in which split-complex numbers exist in the form of (4*4)-matrices $D_L$.

A similar situation holds true for (4*4)-matrices $R_{4R} = R0_{4R} + R1_{4R}$. The set of two matrices $R0_{4R}$ and $R1_{4R}$ is also closed in relation to multiplication; it gives the multiplication table (Figure 35), which is also identical to the multiplication table of split-complex numbers. The set of (4*4)-matrices $D_R = a_1*R0_{4R}+a_3*R1_{4R}$, where $a_1$, $a_3$ are real numbers, represents split-complex numbers in the (4*4)-matrix form (Figure 35). The matrix $R0_{4R}$ plays a role of the real unit in this set of matrices $D_R$. In the case $a_1 \neq a_3$, the matrix $D_R^{-1}$ (Figure 35) is the inverse matrix for $D_R$ since $D_R*D_R^{-1} = D_R^{-1}*D_R = R0_{4R}$.

|   | $R0_{4R}$ | $R1_{4R}$ |
|---|---|---|
| $R0_{4R}$ | $R0_{4R}$ | $R1_{4R}$ |
| $R1_{4R}$ | $R1_{4R}$ | $R0_{4R}$ |

; $D_R = a_1*R0_{4R}+a_3*R1_{4R} = $

| 0 | $a_1$ | 0 | $-a_3$ |
|---|---|---|---|
| 0 | $a_1$ | 0 | $-a_3$ |
| 0 | $-a_3$ | 0 | $a_1$ |
| 0 | $-a_3$ | 0 | $a_1$ |

; $D_R^{-1} = (a_1^2 - a_3^2)^{-1} *$

| 0 | $a_1$ | 0 | $a_3$ |
|---|---|---|---|
| 0 | $a_1$ | 0 | $a_3$ |
| 0 | $a_3$ | 0 | $a_1$ |
| 0 | $a_3$ | 0 | $a_1$ |

Figure 35. The multiplication table of two (4*4)-matrices $R0_{4R}$ and $R1_{4R}$, which is a set of basic elements of split-complex numbers $D_L = a_1*R0_{4L}+a_3*R1_{4L}$, where $a_0$, $a_1$ are real numbers. The matrix $D_L^{-1}$ is the inverse matrix for $D_L$ if $a_1 \neq a_3$.

The initial matrix $R_4$ can be also decomposed in another way: in accordance with the structure of the (4*4)-matrix of dyadic shifts from Figure 1. Figure 36 shows the case of such dyadic-shift decomposition $R_4 = R0_4+R1_4+R2_4+R3_4$ when 4 sparse matrices $R0_4$, $R1_4$, $R2_4$ and $R3_4$ arise ($R0_4$ is identity matrix). The set of these matrices $R0_4$, $R1_4$, $R2_4$ and $R3_4$ is closed in relation to multiplication and it defines the multiplication table on Figure 35. This multiplication table is identical to the multiplication table of 4-dimensional hypercomplex numbers, which are termed as split-quaternions by J.Cockle and which are well known in mathematics and physics (http://en.wikipedia.org/wiki/Split-quaternion). So the genetic matrix $R_4$ is split-quaternion with unit coordinates.

$$\begin{vmatrix} 1 & 1 & 1 & -1 \\ -1 & 1 & -1 & -1 \\ 1 & -1 & 1 & 1 \\ -1 & -1 & -1 & 1 \end{vmatrix} = \begin{vmatrix} 1 & 0 & 0 & 0 \\ 0 & 1 & 0 & 0 \\ 0 & 0 & 1 & 0 \\ 0 & 0 & 0 & 1 \end{vmatrix} + \begin{vmatrix} 0 & 1 & 0 & 0 \\ -1 & 0 & 0 & 0 \\ 0 & 0 & 0 & 1 \\ 0 & 0 & -1 & 0 \end{vmatrix} + \begin{vmatrix} 0 & 0 & 1 & 0 \\ 0 & 0 & 0 & -1 \\ 1 & 0 & 0 & 0 \\ 0 & -1 & 0 & 0 \end{vmatrix} + \begin{vmatrix} 0 & 0 & 0 & -1 \\ 0 & 0 & -1 & 0 \\ 0 & -1 & 0 & 0 \\ -1 & 0 & 0 & 0 \end{vmatrix}$$

|       | $R0_4$ | $R1_4$ | $R2_4$ | $R3_4$ |
|-------|--------|--------|--------|--------|
| $R0_4$ | $R0_4$ | $R1_4$ | $R2_4$ | $R3_4$ |
| $R1_4$ | $R1_4$ | $-R0_4$ | $R3_4$ | $-R2_4$ |
| $R2_4$ | $R2_4$ | $-R3_4$ | $R0_4$ | $-R1_4$ |
| $R3_4$ | $R3_4$ | $R2_4$ | $R1_4$ | $R0_4$ |

Figure 36. Upper row: the dyadic-shift decomposition $R_4 = R0_4+R1_4+R2_4+R3_4$. Bottom row: the multiplication table of the sparse matrices $R0_4$, $R1_4$, $R2_4$ and $R3_4$, which is identical to the multiplication table of split-quaternions by J.Cockle (http://en.wikipedia.org/wiki/Split-quaternion). $R0_4$ is identity matrix for this matrix set and it plays here a role of the real unit in this form of split-quaternions by J.Cockle.

One can numerate 4 columns of the (4*4)-matrix $R_4$ by numbers 0, 1, 2 and 3. Then non-zero columns of the matrix $R_{4L}$ have numeration with even numbers 0 and 2, and non-zero columns of the matrix $R_{4R}$ have numeration with odd numbers 1 and 3. Taking into account the Ancient Chinese tradition to call even numbers as Yin-numbers (or female numbers) and odd numbers as Yang-numbers (or male numbers), these matrices $R_{4L}$ and $R_{4R}$ can be conditionally considered as the Yin half and the Yang half of the whole matrix $R_4$ correspondingly (or the matrices $R_{4L}$ and $R_{4R}$ can be termed as the even part and the odd part of $R_4$, or as the left part and the right part of $R_4$).

Above we have received the interesting result: the sum of two 2-dimensional split-complex numbers $R_{4L}$ and $R_{4R}$ with unit coordinates (they belong to two different matrix types of split-complex numbers) generates the 4-dimensional split-quaternion with unit coordinates. It resembles a situation when a union of Yin and Yang (a union of female and male beginnings, or a fusion of male and female gametes) generates a new organism. Below we will meet with other similar situations concerning to $(2^n*2^n)$-matrices, which are $(2*n)$-dimensional numbers with unit coordinates and which consists of two "complementary" halves (like the matrix $R_4$), each of which is $(2*(n-1))$-dimensional number with unit coordinates. One can name such type of matrices as "matrices with internal complementarities". They resemble in some extend the complementary structure of double helixes of DNA.

Let us return now to the (8*8)-matrix $R_8$ (Figure 32) and demonstrate that it is also the matrix with internal complementarities. Figure 37 shows the matrix $R_8$ as sum of matrices $R_{8L}$ and $R_{8R}$.

| + | + | + | + | + | + | − | − |
|---|---|---|---|---|---|---|---|
| + | + | + | + | + | + | − | − |
| − | − | + | + | − | − | − | − |
| − | − | + | + | − | − | − | − |
| + | + | − | − | + | + | + | + |
| + | + | − | − | + | + | + | + |
| − | − | − | − | − | − | + | + |
| − | − | − | − | − | − | + | + |

=

| + | 0 | + | 0 | + | 0 | − | 0 |
|---|---|---|---|---|---|---|---|
| + | 0 | + | 0 | + | 0 | − | 0 |
| − | 0 | + | 0 | − | 0 | − | 0 |
| − | 0 | + | 0 | − | 0 | − | 0 |
| + | 0 | − | 0 | + | 0 | + | 0 |
| + | 0 | − | 0 | + | 0 | + | 0 |
| − | 0 | − | 0 | − | 0 | + | 0 |
| − | 0 | − | 0 | − | 0 | + | 0 |

+

| 0 | + | 0 | + | 0 | + | 0 | − |
|---|---|---|---|---|---|---|---|
| 0 | + | 0 | + | 0 | + | 0 | − |
| 0 | − | 0 | + | 0 | − | 0 | − |
| 0 | − | 0 | + | 0 | − | 0 | − |
| 0 | + | 0 | − | 0 | + | 0 | + |
| 0 | + | 0 | − | 0 | + | 0 | + |
| 0 | − | 0 | − | 0 | − | 0 | + |
| 0 | − | 0 | − | 0 | − | 0 | + |

Figure 37. The matrix $R_8 = R_{8L}+R_{8R}$. The symbol "+" means "+1" and the symbol "−" means "−1"

Figure 38 shows a decomposition of the matrix $R_{8L}$ (from Figure 37) as a sum of 4 matrices: $R_{8L} = R0_{8L} + R1_{8L} + R2_{8L} + R3_{8L}$. The set of matrices $R0_{8L}$, $R1_{8L}$, $R2_{8L}$ and $R3_{8L}$ is closed in relation to multiplication and it defines the multiplication table, which are identical to the same multiplication table of split-quaternions $R_{8L}$ by J.Cockle. General expression for split-quaternions in this case can be written as $S_L = a_0*R0_{8L} + a_1*R1_{8L} + a_2*R2_{8L} + a_3*R3_{8L}$, where $a_0$, $a_1$, $a_2$, $a_3$ are real numbers. From this point of view, the (8*8)-genomatrix $R_{8L}$ is split-quaternion by Cockle with unit coordinates.

$$\begin{vmatrix} 1 & 0 & 1 & 0 & 1 & 0 & -1 & 0 \\ 1 & 0 & 1 & 0 & 1 & 0 & -1 & 0 \\ -1 & 0 & 1 & 0 & -1 & 0 & -1 & 0 \\ -1 & 0 & 1 & 0 & -1 & 0 & -1 & 0 \\ 1 & 0 & -1 & 0 & 1 & 0 & 1 & 0 \\ 1 & 0 & -1 & 0 & 1 & 0 & 1 & 0 \\ -1 & 0 & -1 & 0 & -1 & 0 & 1 & 0 \\ -1 & 0 & -1 & 0 & -1 & 0 & 1 & 0 \end{vmatrix} = \begin{vmatrix} 1 & 0 & 0 & 0 & 0 & 0 & 0 & 0 \\ 1 & 0 & 0 & 0 & 0 & 0 & 0 & 0 \\ 0 & 0 & 1 & 0 & 0 & 0 & 0 & 0 \\ 0 & 0 & 1 & 0 & 0 & 0 & 0 & 0 \\ 0 & 0 & 0 & 0 & 1 & 0 & 0 & 0 \\ 0 & 0 & 0 & 0 & 1 & 0 & 0 & 0 \\ 0 & 0 & 0 & 0 & 0 & 0 & 1 & 0 \\ 0 & 0 & 0 & 0 & 0 & 0 & 1 & 0 \end{vmatrix} + \begin{vmatrix} 0 & 0 & 1 & 0 & 0 & 0 & 0 & 0 \\ 0 & 0 & 1 & 0 & 0 & 0 & 0 & 0 \\ -1 & 0 & 0 & 0 & 0 & 0 & 0 & 0 \\ -1 & 0 & 0 & 0 & 0 & 0 & 0 & 0 \\ 0 & 0 & 0 & 0 & 0 & 0 & 1 & 0 \\ 0 & 0 & 0 & 0 & 0 & 0 & 1 & 0 \\ 0 & 0 & 0 & 0 & -1 & 0 & 0 & 0 \\ 0 & 0 & 0 & 0 & -1 & 0 & 0 & 0 \end{vmatrix}$$

$$+ \begin{vmatrix} 0 & 0 & 0 & 0 & 1 & 0 & 0 & 0 \\ 0 & 0 & 0 & 0 & 1 & 0 & 0 & 0 \\ 0 & 0 & 0 & 0 & 0 & 0 & -1 & 0 \\ 0 & 0 & 0 & 0 & 0 & 0 & -1 & 0 \\ 1 & 0 & 0 & 0 & 0 & 0 & 0 & 0 \\ 1 & 0 & 0 & 0 & 0 & 0 & 0 & 0 \\ 0 & 0 & -1 & 0 & 0 & 0 & 0 & 0 \\ 0 & 0 & -1 & 0 & 0 & 0 & 0 & 0 \end{vmatrix} + \begin{vmatrix} 0 & 0 & 0 & 0 & 0 & 0 & -1 & 0 \\ 0 & 0 & 0 & 0 & 0 & 0 & -1 & 0 \\ 0 & 0 & 0 & 0 & -1 & 0 & 0 & 0 \\ 0 & 0 & 0 & 0 & -1 & 0 & 0 & 0 \\ 0 & 0 & -1 & 0 & 0 & 0 & 0 & 0 \\ 0 & 0 & -1 & 0 & 0 & 0 & 0 & 0 \\ -1 & 0 & 0 & 0 & 0 & 0 & 0 & 0 \\ -1 & 0 & 0 & 0 & 0 & 0 & 0 & 0 \end{vmatrix}$$

|          | $R0_{8L}$ | $R1_{8L}$  | $R2_{8L}$  | $R3_{8L}$  |
|----------|-----------|------------|------------|------------|
| $R0_{8L}$ | $R0_{8L}$ | $R1_{8L}$  | $R2_{8L}$  | $R3_{8L}$  |
| $R1_{8L}$ | $R1_{8L}$ | $-R0_{8L}$ | $R3_{8L}$  | $-R2_{8L}$ |
| $R2_{8L}$ | $R2_{8L}$ | $-R3_{8L}$ | $R0_{8L}$  | $-R1_{8L}$ |
| $R3_{8L}$ | $R3_{8L}$ | $R2_{8L}$  | $R1_{8L}$  | $R0_{8L}$  |

Figure 38. Upper rows: the decomposition of the matrix $R_{8L}$ (from Figure 37) as sum of 4 matrices: $R_{8L} = R0_{8L} + R1_{8L} + R2_{8L} + R3_{8L}$. Bottom row: the multiplication table of these 4 matrices $R0_{8L}$, $R1_{8L}$, $R2_{8L}$ and $R3_{8L}$, which is identical to the multiplication table of split-quaternions by J.Cockle. $R0_{8L}$ is identity matrix for this matrix set and it plays the role of the real unit here.

The similar situation holds true for the matrix $R_{8R}$ (from Figure 37). Figure 39 shows a decomposition of the matrix $R_{8R}$ as a sum of 4 matrices: $R_{8R} = R0_{8R} + R1_{8R} + R2_{8R} + R3_{8R}$. The set of matrices $R0_{8R}$, $R1_{8R}$, $R2_{8R}$ and $R3_{8R}$ is closed in relation to multiplication and it defines the multiplication table, which are identical to the same multiplication table of split-quaternions by J.Cockle. General expression for split-quaternions in this case can be written as $S_R = a_0*R0_{8R} + a_1*R1_{8R} + a_2*R2_{8R} + a_3*R3_{8R}$, where $a_0$, $a_1$, $a_2$, $a_3$ are real numbers. From this point of view, the (8*8)-genomatrix $R_{8R}$ is split-quaternion with unit coordinates.

$$
\begin{bmatrix}
0 & 1 & 0 & 1 & 0 & 1 & 0 & -1 \\
0 & 1 & 0 & 1 & 0 & 1 & 0 & -1 \\
0 & -1 & 0 & 1 & 0 & -1 & 0 & -1 \\
0 & -1 & 0 & 1 & 0 & -1 & 0 & -1 \\
0 & 1 & 0 & -1 & 0 & 1 & 0 & 1 \\
0 & 1 & 0 & -1 & 0 & 1 & 0 & 1 \\
0 & -1 & 0 & -1 & 0 & -1 & 0 & 1 \\
0 & -1 & 0 & -1 & 0 & -1 & 0 & 1
\end{bmatrix}
=
\begin{bmatrix}
0 & 1 & 0 & 0 & 0 & 0 & 0 & 0 \\
0 & 1 & 0 & 0 & 0 & 0 & 0 & 0 \\
0 & 0 & 0 & 1 & 0 & 0 & 0 & 0 \\
0 & 0 & 0 & 1 & 0 & 0 & 0 & 0 \\
0 & 0 & 0 & 0 & 0 & 1 & 0 & 0 \\
0 & 0 & 0 & 0 & 0 & 1 & 0 & 0 \\
0 & 0 & 0 & 0 & 0 & 0 & 0 & 1 \\
0 & 0 & 0 & 0 & 0 & 0 & 0 & 1
\end{bmatrix}
+
\begin{bmatrix}
0 & 0 & 0 & 1 & 0 & 0 & 0 & 0 \\
0 & 0 & 0 & 1 & 0 & 0 & 0 & 0 \\
0 & -1 & 0 & 0 & 0 & 0 & 0 & 0 \\
0 & -1 & 0 & 0 & 0 & 0 & 0 & 0 \\
0 & 0 & 0 & 0 & 0 & 0 & 0 & 1 \\
0 & 0 & 0 & 0 & 0 & 0 & 0 & 1 \\
0 & 0 & 0 & 0 & 0 & -1 & 0 & 0 \\
0 & 0 & 0 & 0 & 0 & -1 & 0 & 0
\end{bmatrix}
$$

$$
+
\begin{bmatrix}
0 & 0 & 0 & 0 & 0 & 1 & 0 & 0 \\
0 & 0 & 0 & 0 & 0 & 1 & 0 & 0 \\
0 & 0 & 0 & 0 & 0 & 0 & 0 & -1 \\
0 & 0 & 0 & 0 & 0 & 0 & 0 & -1 \\
0 & 1 & 0 & 0 & 0 & 0 & 0 & 0 \\
0 & 1 & 0 & 0 & 0 & 0 & 0 & 0 \\
0 & 0 & 0 & -1 & 0 & 0 & 0 & 0 \\
0 & 0 & 0 & -1 & 0 & 0 & 0 & 0
\end{bmatrix}
+
\begin{bmatrix}
0 & 0 & 0 & 0 & 0 & 0 & 0 & -1 \\
0 & 0 & 0 & 0 & 0 & 0 & 0 & -1 \\
0 & 0 & 0 & 0 & 0 & -1 & 0 & 0 \\
0 & 0 & 0 & 0 & 0 & -1 & 0 & 0 \\
0 & 0 & 0 & -1 & 0 & 0 & 0 & 0 \\
0 & 0 & 0 & -1 & 0 & 0 & 0 & 0 \\
0 & -1 & 0 & 0 & 0 & 0 & 0 & 0 \\
0 & -1 & 0 & 0 & 0 & 0 & 0 & 0
\end{bmatrix}
$$

|           | $R0_{8R}$ | $R1_{8R}$  | $R2_{8R}$  | $R3_{8L}$  |
|-----------|-----------|------------|------------|------------|
| $R0_{8R}$ | $R0_{8R}$ | $R1_{8R}$  | $R2_{8R}$  | $R3_{8R}$  |
| $R1_{8R}$ | $R1_{8R}$ | $-R0_{8R}$ | $R3_{8R}$  | $-R2_{8R}$ |
| $R2_{8R}$ | $R2_{8R}$ | $-R3_{8R}$ | $R0_{8R}$  | $-R1_{8R}$ |
| $R3_{8R}$ | $R3_{8R}$ | $R2_{8R}$  | $R1_{8R}$  | $R0_{8R}$  |

Figure 39. Upper rows: the decomposition of the matrix $R_{8R}$ (from Figure 37) as sum of 4 matrices: $R_{8R} = R0_{8R} + R1_{8R} + R2_{8R} + R3_{8L}$. Bottom row: the multiplication table of these 4 matrices $R0_{8R}$, $R1_{8R}$, $R2_{8R}$ and $R3_{8L}$, which is identical to the multiplication table of split-quaternions by J.Cockle. $R0_{8R}$ is identity matrix for this matrix set and it plays the role of the real unit here.

The initial (8*8)-matrix $R_8$ can be also decomposed in another way, or more precisely in accordance with the structure of the (8*8)-matrix of dyadic shifts from Figure 1. Figure 40 shows the case of such dyadic-shift decomposition $R_8 = R0_8+R1_8+R2_8+R3_8+R4_8+R5_8+R6_8+R7_8$, when 8 sparse matrices $R0_8$, $R1_8$, $R2_8$, $R3_8$, $R4_8$, $R5_8$, $R6_8$, $R7_8$ arise ($R0_8$ is identity matrix for this matrix set). The set $R0_8$, $R1_8$, $R2_8$, $R3_8$, $R4_8$, $R5_8$, $R6_8$, $R7_8$ is closed in relation to multiplication and it defines the multiplication table on Figure 40. This multiplication table is identical to the multiplication table of 8-dimensional hypercomplex numbers, which are termed as bi-split-quaternions by J.Cockle (or split-quaternions over the field of complex numbers). General expression for bi-split-quaternions in this case can be written as $S_8 = a_0*R0_8+a_1*R1_8+a_2*R2_8+a_3*R3_8+a_4*R4_8+a_5*R5_8+a_6*R6_8+a_7*R7_8$, where $a_0$, $a_1$, $a_2$, $a_3$, $a_4$, $a_5$, $a_6$, $a_7$ are real numbers. From this point of view, the (8*8)-genomatrix $R_8$ is bi-split-quaternion with unit coordinates.

$R_8 = R0_8+R1_8+R2_8+R3_8+R4_8+R5_8+R6_8+R7_8 =$

$$
\begin{bmatrix}
1 & 0 & 0 & 0 & 0 & 0 & 0 & 0 \\
0 & 1 & 0 & 0 & 0 & 0 & 0 & 0 \\
0 & 0 & 1 & 0 & 0 & 0 & 0 & 0 \\
0 & 0 & 0 & 1 & 0 & 0 & 0 & 0 \\
0 & 0 & 0 & 0 & 1 & 0 & 0 & 0 \\
0 & 0 & 0 & 0 & 0 & 1 & 0 & 0 \\
0 & 0 & 0 & 0 & 0 & 0 & 1 & 0 \\
0 & 0 & 0 & 0 & 0 & 0 & 0 & 1
\end{bmatrix}
+
\begin{bmatrix}
0 & 1 & 0 & 0 & 0 & 0 & 0 & 0 \\
1 & 0 & 0 & 0 & 0 & 0 & 0 & 0 \\
0 & 0 & 0 & 1 & 0 & 0 & 0 & 0 \\
0 & 0 & 1 & 0 & 0 & 0 & 0 & 0 \\
0 & 0 & 0 & 0 & 0 & 1 & 0 & 0 \\
0 & 0 & 0 & 0 & 1 & 0 & 0 & 0 \\
0 & 0 & 0 & 0 & 0 & 0 & 0 & 1 \\
0 & 0 & 0 & 0 & 0 & 0 & 1 & 0
\end{bmatrix}
+
\begin{bmatrix}
0 & 0 & 1 & 0 & 0 & 0 & 0 & 0 \\
0 & 0 & 0 & 1 & 0 & 0 & 0 & 0 \\
-1 & 0 & 0 & 0 & 0 & 0 & 0 & 0 \\
0 & -1 & 0 & 0 & 0 & 0 & 0 & 0 \\
0 & 0 & 0 & 0 & 0 & 0 & 1 & 0 \\
0 & 0 & 0 & 0 & 0 & 0 & 0 & 1 \\
0 & 0 & 0 & 0 & -1 & 0 & 0 & 0 \\
0 & 0 & 0 & 0 & 0 & -1 & 0 & 0
\end{bmatrix}
+
\begin{bmatrix}
0 & 0 & 0 & 1 & 0 & 0 & 0 & 0 \\
0 & 0 & 1 & 0 & 0 & 0 & 0 & 0 \\
0 & -1 & 0 & 0 & 0 & 0 & 0 & 0 \\
-1 & 0 & 0 & 0 & 0 & 0 & 0 & 0 \\
0 & 0 & 0 & 0 & 0 & 0 & 0 & 1 \\
0 & 0 & 0 & 0 & 0 & 0 & 1 & 0 \\
0 & 0 & 0 & 0 & 0 & -1 & 0 & 0 \\
0 & 0 & 0 & 0 & -1 & 0 & 0 & 0
\end{bmatrix}
+
$$

$$\begin{pmatrix} 0 & 0 & 0 & 0 & 1 & 0 & 0 & 0 \\ 0 & 0 & 0 & 0 & 0 & 1 & 0 & 0 \\ 0 & 0 & 0 & 0 & 0 & 0 & -1 & 0 \\ 0 & 0 & 0 & 0 & 0 & 0 & 0 & -1 \\ 1 & 0 & 0 & 0 & 0 & 0 & 0 & 0 \\ 0 & 1 & 0 & 0 & 0 & 0 & 0 & 0 \\ 0 & 0 & -1 & 0 & 0 & 0 & 0 & 0 \\ 0 & 0 & 0 & -1 & 0 & 0 & 0 & 0 \end{pmatrix} + \begin{pmatrix} 0 & 0 & 0 & 0 & 0 & 1 & 0 & 0 \\ 0 & 0 & 0 & 0 & 1 & 0 & 0 & 0 \\ 0 & 0 & 0 & 0 & 0 & 0 & 0 & -1 \\ 0 & 0 & 0 & 0 & 0 & 0 & -1 & 0 \\ 0 & 1 & 0 & 0 & 0 & 0 & 0 & 0 \\ 1 & 0 & 0 & 0 & 0 & 0 & 0 & 0 \\ 0 & 0 & 0 & -1 & 0 & 0 & 0 & 0 \\ 0 & 0 & -1 & 0 & 0 & 0 & 0 & 0 \end{pmatrix} + \begin{pmatrix} 0 & 0 & 0 & 0 & 0 & 0 & -1 & 0 \\ 0 & 0 & 0 & 0 & 0 & 0 & 0 & -1 \\ 0 & 0 & 0 & 0 & -1 & 0 & 0 & 0 \\ 0 & 0 & 0 & 0 & 0 & -1 & 0 & 0 \\ 0 & 0 & -1 & 0 & 0 & 0 & 0 & 0 \\ 0 & 0 & 0 & -1 & 0 & 0 & 0 & 0 \\ -1 & 0 & 0 & 0 & 0 & 0 & 0 & 0 \\ 0 & -1 & 0 & 0 & 0 & 0 & 0 & 0 \end{pmatrix} + \begin{pmatrix} 0 & 0 & 0 & 0 & 0 & 0 & 0 & -1 \\ 0 & 0 & 0 & 0 & 0 & 0 & -1 & 0 \\ 0 & 0 & 0 & 0 & 0 & -1 & 0 & 0 \\ 0 & 0 & 0 & 0 & -1 & 0 & 0 & 0 \\ 0 & 0 & 0 & -1 & 0 & 0 & 0 & 0 \\ 0 & 0 & -1 & 0 & 0 & 0 & 0 & 0 \\ 0 & -1 & 0 & 0 & 0 & 0 & 0 & 0 \\ -1 & 0 & 0 & 0 & 0 & 0 & 0 & 0 \end{pmatrix}$$

|        | $R0_8$ | $R1_8$ | $R2_8$ | $R3_8$ | $R4_8$ | $R5_8$ | $R6_8$ | $R7_8$ |
|--------|--------|--------|--------|--------|--------|--------|--------|--------|
| $R0_8$ | $R0_8$ | $R1_8$ | $R2_8$ | $R3_8$ | $R4_8$ | $R5_8$ | $R6_8$ | $R7_8$ |
| $R1_8$ | $R1_8$ | $R0_8$ | $R3_8$ | $R2_8$ | $R5_8$ | $R4_8$ | $R7_8$ | $R6_8$ |
| $R2_8$ | $R2_8$ | $R3_8$ | $-R0_8$ | $-R1_8$ | $R6_8$ | $R7_8$ | $-R4_8$ | $-R5_8$ |
| $R3_8$ | $R3_8$ | $R2_8$ | $-R1_8$ | $-R0_8$ | $R7_8$ | $R6_8$ | $-R5_8$ | $-R4_8$ |
| $R4_8$ | $R4_8$ | $R5_8$ | $-R6_8$ | $-R7_8$ | $R0_8$ | $R1_8$ | $-R2_8$ | $-R3_8$ |
| $R5_8$ | $R5_8$ | $R4_8$ | $-R7_8$ | $-R6_8$ | $R1_8$ | $R0_8$ | $-R3_8$ | $-R2_8$ |
| $R6_8$ | $R6_8$ | $R7_8$ | $R4_8$ | $R5_8$ | $R2_8$ | $R3_8$ | $R0_8$ | $R1_8$ |
| $R7_8$ | $R7_8$ | $R6_8$ | $R5_8$ | $R4_8$ | $R3_8$ | $R2_8$ | $R1_8$ | $R0_8$ |

Figure 40. Upper rows: the decomposition of the matrix $R_8$ (from Figure 32) as sum of 8 matrices: $R_8 = R0_8+R1_8+R2_8+R3_8+R4_8+R5_8+R6_8+R7_8$. Bottom row: the multiplication table of these 8 matrices $R0_8$, $R1_8$, $R2_8$, $R3_8$, $R4_8$, $R5_8$, $R6_8$ and $R7_8$, which is identical to the multiplication table of bi-split-quaternions by J.Cockle (or split-quaternions over the field of complex numbers). $R0_8$ is identity matrix for this matrix set and it plays the role of the real unit here.

Here for the (8*8)-genomatrix $R_8$ we have received the interesting result: the sum of two different 4-dimensional split-quaternions by Cockle with unit coordinates (they belong to two different matrix types of split-quaternion numbers) generates the 8-dimensional bi-split-quaternion with unit coordinates. This result resembles the above-described result about the sum of 2-dimensional split-complex numbers with unit coordinates, which generates the 4-dimensional split-quaternion with unit coordinates (Figures 34-36). It resembles also a situation when a union of Yin and Yang (a union of male and female beginnings, or a fusion of male and female gametes) generates a new organism.

Now let us pay attention to Hadamard genomatrices $H_4$ and $H_8$ (Figure 41), which belong to the second important type of genetic matrices and which are closely connected with Rademacher genomatrices $R_4$ and $R_8$ from Figures 31 and 32. This connection is provided on the special T/U algorithm, which is based on objective properties of the 4-letter genetic alphabet A, C, G, T and which is described in our previous publications [Petoukhov, 2008a,b, 2011a,b, 2012; Petoukhov, He, 2010]. For the T/U-algorithm, phenomenological facts are essential that the letter T in DNA (and correspondingly the letter U in RNA) is a very special letter in the 4-letter alphabet of nitrogenous bases in the following two aspects:

• Each of three nitrogenous bases A, C, G has one amino-group NH2, but the fourth basis T/U has not it. From the viewpoint of existence of the amino-group (which is very important for genetic functions) the letters A, C, G are identical to each other and the letter T is opposite to them;

• The letter T is a single letter in DNA, which is replaced in RNA by another letter U.

This uniqueness of the letter T can be utilized in genetic computers of organisms. By definition the genetic T/U-algorithm contains two steps: 1) on the first step, each of duplets or triplets in the black-and-white genomatrices (for example, in the genomatrix [C T; A G]$^{(3)}$ on Figure 31) should change its own color into opposite color each time when the letter T stands in an odd position (in the first position inside the duplet or in the first or third position inside the triplet); 2) on the second step, black triplets and white triples are interpreted as entries "+1" and "-1" correspondingly. For example, the white triplet TTA (see Figure 31) should become the black triplet (and its matrix cell should be marked by black color) because of the letter T in its first position; for this reason the triplet TTA is interpreted finally as "+1". Or the white triplet TTT should not change its color because of the letter T in its first and third positions (the color of this triplet is changed twice according to the described algorithm); for this reason the triplet TTT is interpreted finally as "-1". The triplet ACG does not change its color because the letter T is absent in this triplet at all.

By means of the genetic T/U-algorithm, the black-and-white mosaic of the genomatrices [C T; A G]$^{(2)}$ and [C T; A G]$^{(3)}$ (Figure 31) receive new black-and-white mosaics (Figure 40), which are identical to mosaics of a disposition of entries +1 and -1 in Hadamard (4*4)- and (8*8)-matrices $H_4$ and $H_8$ (Figure 41). Hadamard matrices have only two types of entries: +1 and -1. One can remind that Hadamard matrices are used widely in modern digital communication, quantum mechanics and many other fields. They are also connected with the genetic coding system [Petoukhov, 2008a,b, 2011a,b, 2012; Petoukhov, He, 2010].

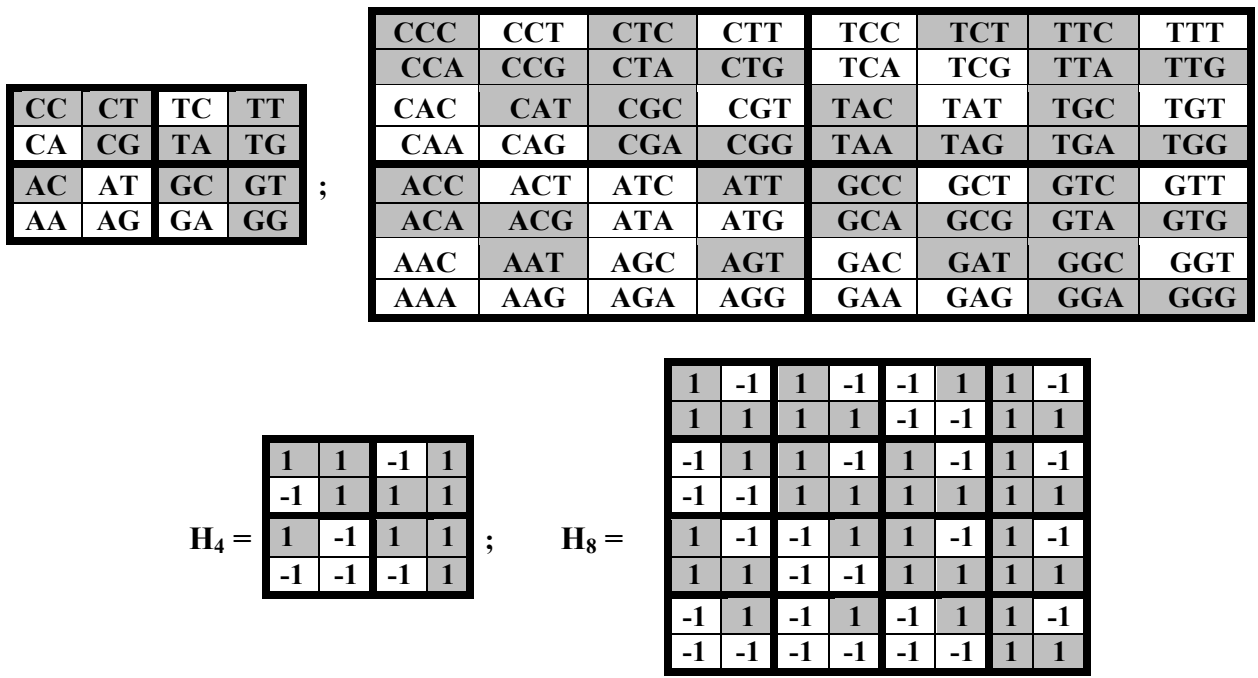

Figure 41. Upper row: genetic matrices [C T; A G]$^{(2)}$ and [C T; A G]$^{(3)}$ with new mosaics, which are received in the result of transformation of their mosaics on Figure 31 by means of the genetic T/U-algorithm. Bottom row: corresponding Hadamard matrices $H_4$ and $H_8$ with the same mosaics.

Let us initially analyze the matrix $H_4$. It is a sum of two matrices $H_{4L}$ and $H_{4R}$ (Figure 42).

$$H_4 = H_{4L} + H_{4R} = \begin{array}{|c|c|c|c|} \hline 1 & 0 & -1 & 0 \\ \hline -1 & 0 & 1 & 0 \\ \hline 1 & 0 & 1 & 0 \\ \hline -1 & 0 & -1 & 0 \\ \hline \end{array} + \begin{array}{|c|c|c|c|} \hline 0 & 1 & 0 & 1 \\ \hline 0 & 1 & 0 & 1 \\ \hline 0 & -1 & 0 & 1 \\ \hline 0 & -1 & 0 & 1 \\ \hline \end{array}$$ , where

$$H_{4L} = H0_{4L} + H1_{4L} = \begin{bmatrix} 1 & 0 & 0 & 0 \\ -1 & 0 & 0 & 0 \\ 0 & 0 & 1 & 0 \\ 0 & 0 & -1 & 0 \end{bmatrix} + \begin{bmatrix} 0 & 0 & -1 & 0 \\ 0 & 0 & 1 & 0 \\ 1 & 0 & 0 & 0 \\ -1 & 0 & 0 & 0 \end{bmatrix},$$

$$H_{4R} = H0_{4R} + H1_{4R} = \begin{bmatrix} 0 & 1 & 0 & 0 \\ 0 & 1 & 0 & 0 \\ 0 & 0 & 0 & 1 \\ 0 & 0 & 0 & 1 \end{bmatrix} + \begin{bmatrix} 0 & 0 & 0 & 1 \\ 0 & 0 & 0 & 1 \\ 0 & -1 & 0 & 0 \\ 0 & -1 & 0 & 0 \end{bmatrix}$$

Figure 42. Upper row: the representation of the matrix $H_4$ as sum of matrices $H_{4L}$ and $H_{4R}$. Other rows: representations of matrices $H_{4L}$ and $H_{4R}$ as sums of matrices $H0_{4L}$, $H1_{4L}$, $H0_{4R}$ and $H1_{4R}$.

It is unexpected but the set of two (4*4)-matrices $H0_{4L}$ and $H1_{4L}$ is also closed in relation to multiplication and it defines the multiplication table of these matrices (Figure 43), which is identical to the multiplication table of complex numbers (http://en.wikipedia.org/wiki/Complex_number).

|        | $H0_{4L}$ | $H1_{4L}$ |
|--------|-----------|-----------|
| $H0_{4L}$ | $H0_{4L}$ | $H1_{4L}$ |
| $H1_{4L}$ | $H1_{4L}$ | $-H0_{4L}$ |

; $C_L = a_0* H0_{4L} + a_2*H1_{4L} = \begin{bmatrix} a_0 & 0 & -a_2 & 0 \\ -a_0 & 0 & a_2 & 0 \\ a_2 & 0 & a_0 & 0 \\ -a_2 & 0 & -a_0 & 0 \end{bmatrix}$ ; $C_L^{-1} = (a_0^2 + a_2^2)^{-1} * \begin{bmatrix} a_0 & 0 & a_2 & 0 \\ -a_0 & 0 & -a_2 & 0 \\ -a_2 & 0 & a_0 & 0 \\ a_2 & 0 & -a_0 & 0 \end{bmatrix}$

Figure 43. The multiplication table of two (4*4)-matrices $H0_{4L}$ and $H1_{4L}$ (Figure 42), which is a set of two basic elements of complex numbers $C_L = a_0*H0_{4L} + a_2*H1_{4L}$, where $a_0$, $a_2$ are real numbers. The matrix $C_L^{-1}$ is the inverse matrix for $C_L$.

The set of (4*4)-matrices $C_L = a_0*H0_{4L} + a_2*H1_{4L}$, where $a_0$, $a_2$ are real numbers, represents complex numbers in the special (4*4)-matrix form (Figure 42). The classical identity matrix E=[1 0 0 0; 0 1 0 0; 0 0 1 0; 0 0 0 1] is absent in the set of matrices $C_L$, where the matrix $H0_{4L}$ plays a role of the real unit (identity matrix for this set). The matrix $C_L^{-1}$ (Figure 42) is the inverse matrix for $C_L$ since $C_L*C_L^{-1} = C_L^{-1}*C_L = H0_{4L}$. From this point of view, the genetic matrix $H_{4L}$ is complex number with unit coordinates.

One should note that the (2*2)-matrix $[a_0\ a_2;\ -a_2\ a_0]$ is usually used for a matrix representation of complex numbers. In the case of genetic matrices, we reveal that 4-dimensional spaces can contain 2-parametric subspaces, in which complex numbers exist in the form of (4*4)-matrices $C_L$.

A similar situation holds true for (4*4)-matrices $H_{4R} = H0_{4R} + H1_{4R}$ (Figure 42). The set of two matrices $H0_{4R}$ and $H1_{4R}$ is also closed in relation to multiplication; it gives the multiplication table (Figure 44), which is also identical to the multiplication table of complex numbers. The set of (4*4)-matrices $C_R = a_1*H0_{4R} + a_3*H1_{4R}$, where $a_1$, $a_3$ are real numbers, represents complex numbers in the (4*4)-matrix form (Figure 44). The matrix $H0_{4R}$ plays a role of the real unit in this set of matrices $C_R$. The matrix $C_R^{-1}$ (Figure 44) is the inverse matrix for $C_R$ since $C_R*C_R^{-1} = C_R^{-1}*C_R = H0_{4R}$. From this point of view, the genetic matrix $H_{4R}$ is complex number with unit coordinates.

|        | $H0_{4R}$ | $H1_{4R}$ |
|--------|-----------|-----------|
| $H0_{4R}$ | $H0_{4R}$ | $H1_{4R}$ |
| $H1_{4R}$ | $H1_{4R}$ | $-H0_{4R}$ |

; $C_R = a_1* H0_{4R} + a_3*H1_{4R} = \begin{bmatrix} 0 & a_1 & 0 & a_3 \\ 0 & a_1 & 0 & a_3 \\ 0 & -a_3 & 0 & a_1 \\ 0 & -a_3 & 0 & a_1 \end{bmatrix}$ ; $C_R^{-1} = (a_1^2 + a_3^2)^{-1} * \begin{bmatrix} 0 & a_1 & 0 & -a_3 \\ 0 & a_1 & 0 & -a_3 \\ 0 & a_3 & 0 & a_1 \\ 0 & a_3 & 0 & a_1 \end{bmatrix}$

Figure 44. The multiplication table of two (4*4)-matrices $H0_{4R}$ and $H1_{4R}$ (Figure 42), which is a set of two basic elements of complex numbers $C_R = a_1*H0_{4L}+a_3*H1_{4R}$, where $a_1$, $a_3$ are real numbers. The matrix $C_R^{-1}$ is the inverse matrix for $C_R$.

The initial matrix $H_4$ can be also decomposed in another way: in accordance with the structure of the (4*4)-matrix of dyadic shifts from Figure 1. Figure 45 shows the case of such dyadic-shift decomposition $H_4 = H0_4+H1_4+H2_4+H3_4$ when 4 sparse matrices $H0_4$, $H1_4$, $H2_4$ and $H3_4$ arise ($H0_4$ is identity matrix). The set of these matrices $H0_4$, $H1_4$, $H2_4$ and $H3_4$ is closed in relation to multiplication and it defines the multiplication table on Figure 45. This multiplication table is identical to the multiplication table of 4-dimensional quaternions by W.Hamilton, well known in mathematics and physics (http://en.wikipedia.org/wiki/Quaternion). So the genetic matrix $H_4$ is quaternion by Hamilton with unit coordinates.

$$H_4 = H0_4+H1_4+H2_4+H3_4 = \begin{vmatrix} 1 & 0 & 0 & 0 \\ 0 & 1 & 0 & 0 \\ 0 & 0 & 1 & 0 \\ 0 & 0 & 0 & 1 \end{vmatrix} + \begin{vmatrix} 0 & 1 & 0 & 0 \\ -1 & 0 & 0 & 0 \\ 0 & 0 & 0 & 1 \\ 0 & 0 & -1 & 0 \end{vmatrix} + \begin{vmatrix} 0 & 0 & -1 & 0 \\ 0 & 0 & 0 & 1 \\ 1 & 0 & 0 & 0 \\ 0 & -1 & 0 & 0 \end{vmatrix} + \begin{vmatrix} 0 & 0 & 0 & 1 \\ 0 & 0 & 1 & 0 \\ 0 & -1 & 0 & 0 \\ -1 & 0 & 0 & 0 \end{vmatrix}$$

|        | $H0_4$ | $H1_4$ | $H2_4$ | $H3_4$ |
|--------|--------|--------|--------|--------|
| $H0_4$ | $H0_4$ | $H1_4$ | $H2_4$ | $H3_4$ |
| $H1_4$ | $H1_4$ | $-H0_4$ | $H3_4$ | $-H2_4$ |
| $H2_4$ | $H2_4$ | $-H3_4$ | $-H0_4$ | $H1_4$ |
| $H3_4$ | $H3_4$ | $H2_4$ | $-H1_4$ | $-H0_4$ |

Figure 45. Upper row: the dyadic-shift decomposition of the genetic matrix $H_4 = H0_4+H1_4+H2_4+H3_4$. Bottom row: the multiplication table of the sparse matrices $H0_4$, $H1_4$, $H2_4$ and $H3_4$, which is identical to multiplication table of quaternions by Hamilton ($H0_4$ is identity matrix).

Let us return now to the (8*8)-matrix $H_8$ (Figure 41) and demonstrate that it is also the matrix with internal complementarities. Figure 46 shows the matrix $H_8$ as sum of matrices $H_{8L}$ and $H_{8R}$.

| 1 | -1 | 1 | -1 | -1 | 1 | 1 | -1 |   | 1 | 0 | 1 | 0 | -1 | 0 | 1 | 0 |   | 0 | -1 | 0 | -1 | 0 | 1 | 0 | -1 |
|---|----|---|----|----|---|---|----|---|---|---|---|---|----|---|---|---|---|---|----|---|----|---|---|---|----|
| 1 | 1 | 1 | 1 | -1 | -1 | 1 | 1 |   | 1 | 0 | 1 | 0 | -1 | 0 | 1 | 0 |   | 0 | 1 | 0 | 1 | 0 | -1 | 0 | 1 |
| -1 | 1 | 1 | -1 | 1 | -1 | 1 | -1 |   | -1 | 0 | 1 | 0 | 1 | 0 | 1 | 0 |   | 0 | 1 | 0 | -1 | 0 | -1 | 0 | -1 |
| -1 | -1 | 1 | 1 | 1 | 1 | 1 | 1 | = | -1 | 0 | 1 | 0 | 1 | 0 | 1 | 0 | + | 0 | -1 | 0 | 1 | 0 | 1 | 0 | 1 |
| 1 | -1 | -1 | 1 | 1 | -1 | 1 | -1 |   | 1 | 0 | -1 | 0 | 1 | 0 | 1 | 0 |   | 0 | -1 | 0 | 1 | 0 | -1 | 0 | -1 |
| 1 | 1 | -1 | -1 | 1 | 1 | 1 | 1 |   | 1 | 0 | -1 | 0 | 1 | 0 | 1 | 0 |   | 0 | 1 | 0 | -1 | 0 | 1 | 0 | 1 |
| -1 | 1 | -1 | 1 | -1 | 1 | 1 | -1 |   | -1 | 0 | -1 | 0 | -1 | 0 | 1 | 0 |   | 0 | 1 | 0 | 1 | 0 | 1 | 0 | -1 |
| -1 | -1 | -1 | -1 | -1 | -1 | 1 | 1 |   | -1 | 0 | -1 | 0 | -1 | 0 | 1 | 0 |   | 0 | -1 | 0 | -1 | 0 | -1 | 0 | 1 |

Figure 46. The matrix $H_8 = H_{8L}+H_{8R}$ is one of matrices with internal complementarities

Figure 47 shows a decomposition of the matrix $H_{8L}$ (from Figure 46) as a sum of 4 matrices: $H_{8L} = H0_{8L} + H1_{8L} + H2_{8L} + H3_{8L}$. The set of matrices $H0_{8L}$, $H1_{8L}$, $H2_{8L}$ and $H3_{8L}$ is closed in relation to multiplication and it defines the multiplication table, which are identical to the same multiplication table of quaternions by Hamilton. General expression for quaternions in this case can be written as $Q_L = a_0*H0_{8L} + a_1*H1_{8L} + a_2*H2_{8L} + a_3*H3_{8L}$, where $a_0$, $a_1$, $a_2$, $a_3$ are real numbers. From this point of view, the (8*8)-genomatrix $H_{8L}$ is 4-dimensional quaternion by Hamilton with unit coordinates.

$$H_{8L} = H0_{8L} + H1_{8L} + H2_{8L} + H3_{8L} =$$

$$
\begin{vmatrix} 1\ 0\ 1\ 0\ -1\ 0\ 1\ 0 \\ 1\ 0\ 1\ 0\ -1\ 0\ 1\ 0 \\ -1\ 0\ 1\ 0\ 1\ 0\ 1\ 0 \\ -1\ 0\ 1\ 0\ 1\ 0\ 1\ 0 \\ 1\ 0\ -1\ 0\ 1\ 0\ 1\ 0 \\ 1\ 0\ -1\ 0\ 1\ 0\ 1\ 0 \\ -1\ 0\ -1\ 0\ -1\ 0\ 1\ 0 \\ -1\ 0\ -1\ 0\ -1\ 0\ 1\ 0 \end{vmatrix} = \begin{vmatrix} 1\ 0\ 0\ 0\ 0\ 0\ 0\ 0 \\ 1\ 0\ 0\ 0\ 0\ 0\ 0\ 0 \\ 0\ 0\ 1\ 0\ 0\ 0\ 0\ 0 \\ 0\ 0\ 1\ 0\ 0\ 0\ 0\ 0 \\ 0\ 0\ 0\ 0\ 1\ 0\ 0\ 0 \\ 0\ 0\ 0\ 0\ 1\ 0\ 0\ 0 \\ 0\ 0\ 0\ 0\ 0\ 0\ 1\ 0 \\ 0\ 0\ 0\ 0\ 0\ 0\ 1\ 0 \end{vmatrix} + \begin{vmatrix} 0\ 0\ 1\ 0\ 0\ 0\ 0\ 0 \\ 0\ 0\ 1\ 0\ 0\ 0\ 0\ 0 \\ -1\ 0\ 0\ 0\ 0\ 0\ 0\ 0 \\ -1\ 0\ 0\ 0\ 0\ 0\ 0\ 0 \\ 0\ 0\ 0\ 0\ 0\ 0\ 1\ 0 \\ 0\ 0\ 0\ 0\ 0\ 0\ 1\ 0 \\ 0\ 0\ 0\ 0\ -1\ 0\ 0\ 0 \\ 0\ 0\ 0\ 0\ -1\ 0\ 0\ 0 \end{vmatrix}
$$

$$
+ \begin{vmatrix} 0\ 0\ 0\ 0\ -1\ 0\ 0\ 0 \\ 0\ 0\ 0\ 0\ -1\ 0\ 0\ 0 \\ 0\ 0\ 0\ 0\ 0\ 0\ 1\ 0 \\ 0\ 0\ 0\ 0\ 0\ 0\ 1\ 0 \\ 1\ 0\ 0\ 0\ 0\ 0\ 0\ 0 \\ 1\ 0\ 0\ 0\ 0\ 0\ 0\ 0 \\ 0\ 0\ -1\ 0\ 0\ 0\ 0\ 0 \\ 0\ 0\ -1\ 0\ 0\ 0\ 0\ 0 \end{vmatrix} + \begin{vmatrix} 0\ 0\ 0\ 0\ 0\ 0\ 1\ 0 \\ 0\ 0\ 0\ 0\ 0\ 0\ 1\ 0 \\ 0\ 0\ 0\ 0\ 1\ 0\ 0\ 0 \\ 0\ 0\ 0\ 0\ 1\ 0\ 0\ 0 \\ 0\ 0\ -1\ 0\ 0\ 0\ 0\ 0 \\ 0\ 0\ -1\ 0\ 0\ 0\ 0\ 0 \\ -1\ 0\ 0\ 0\ 0\ 0\ 0\ 0 \\ -1\ 0\ 0\ 0\ 0\ 0\ 0\ 0 \end{vmatrix}
$$

|         | $H0_{8L}$ | $H1_{8L}$ | $H2_{8L}$ | $H3_{8L}$ |
|---------|-----------|-----------|-----------|-----------|
| $H0_{8L}$ | $H0_{8L}$ | $H1_{8L}$ | $H2_{8L}$ | $H3_{8L}$ |
| $H1_{8L}$ | $H1_{8L}$ | $-H0_{8L}$ | $H3_{8L}$ | $-H2_{8L}$ |
| $H2_{8L}$ | $H2_{8L}$ | $-H3_{8L}$ | $-H0_{8L}$ | $H1_{8L}$ |
| $H3_{8L}$ | $H3_{8L}$ | $H2_{8L}$ | $-H1_{8L}$ | $-H0_{8L}$ |

Figure 47. Upper rows: the decomposition of the matrix $H_{8L}$ (from Figure 46) as sum of 4 matrices: $H_{8L} = H0_{8L} + H1_{8L} + H2_{8L} + H3_{8L}$. Bottom row: the multiplication table of these 4 matrices $H0_{8L}$, $H1_{8L}$, $H2_{8L}$ and $H3_{8L}$, which is identical to the multiplication table of quaternions by Hamilton. $H0_{8L}$ is identity matrix for this matrix set and it plays the role of the real unit here.

The similar situation holds true for the matrix $H_{8R}$ (from Figure 46). Figure 48 shows a decomposition of the matrix $H_{8R}$ as a sum of 4 matrices: $H_{8R} = H0_{8R} + H1_{8R} + H2_{8R} + H3_{8R}$. The set of matrices $H0_{8R}$, $H1_{8R}$, $H2_{8R}$ and $H3_{8R}$ is closed in relation to multiplication and it defines the multiplication table, which are identical to the same multiplication table of quaternions by Hamilton. General expression for quaternions in this case can be written as $Q_R = a_0*H0_{8R} + a_2*H1_{8R} + a_4*H2_{8R} + a_6*H3_{8R}$, where $a_0, a_2, a_4, a_6$ are real numbers. From this point of view, the (8*8)-genomatrix $H_{8R}$ is quaternion by Hamilton with unit coordinates.

$H_{8R} = H0_{8R} + H1_{8R} + H2_{8R} + H3_{8R} =$

$$
\begin{vmatrix} 0\ -1\ 0\ -1\ 0\ 1\ 0\ -1 \\ 0\ 1\ 0\ 1\ 0\ -1\ 0\ 1 \\ 0\ 1\ 0\ -1\ 0\ -1\ 0\ -1 \\ 0\ -1\ 0\ 1\ 0\ 1\ 0\ 1 \\ 0\ -1\ 0\ 1\ 0\ -1\ 0\ -1 \\ 0\ 1\ 0\ -1\ 0\ 1\ 0\ 1 \\ 0\ 1\ 0\ 1\ 0\ 1\ 0\ -1 \\ 0\ -1\ 0\ -1\ 0\ -1\ 0\ 1 \end{vmatrix} = \begin{vmatrix} 0\ -1\ 0\ 0\ 0\ 0\ 0\ 0 \\ 0\ 1\ 0\ 0\ 0\ 0\ 0\ 0 \\ 0\ 0\ 0\ -1\ 0\ 0\ 0\ 0 \\ 0\ 0\ 0\ 1\ 0\ 0\ 0\ 0 \\ 0\ 0\ 0\ 0\ 0\ -1\ 0\ 0 \\ 0\ 0\ 0\ 0\ 0\ 1\ 0\ 0 \\ 0\ 0\ 0\ 0\ 0\ 0\ 0\ -1 \\ 0\ 0\ 0\ 0\ 0\ 0\ 0\ 1 \end{vmatrix} + \begin{vmatrix} 0\ 0\ 0\ -1\ 0\ 0\ 0\ 0 \\ 0\ 0\ 0\ 1\ 0\ 0\ 0\ 0 \\ 0\ 1\ 0\ 0\ 0\ 0\ 0\ 0 \\ 0\ -1\ 0\ 0\ 0\ 0\ 0\ 0 \\ 0\ 0\ 0\ 0\ 0\ 0\ 0\ -1 \\ 0\ 0\ 0\ 0\ 0\ 0\ 0\ 1 \\ 0\ 0\ 0\ 0\ 0\ 0\ 1\ 0 \\ 0\ 0\ 0\ 0\ 0\ -1\ 0\ 0 \end{vmatrix}
$$

$$
+\begin{bmatrix} 0 & 0 & 0 & 0 & 0 & 1 & 0 & 0 \\ 0 & 0 & 0 & 0 & 0 & -1 & 0 & 0 \\ 0 & 0 & 0 & 0 & 0 & 0 & 0 & -1 \\ 0 & 0 & 0 & 0 & 0 & 0 & 0 & 1 \\ 0 & -1 & 0 & 0 & 0 & 0 & 0 & 0 \\ 0 & 1 & 0 & 0 & 0 & 0 & 0 & 0 \\ 0 & 0 & 0 & 1 & 0 & 0 & 0 & 0 \\ 0 & 0 & 0 & -1 & 0 & 0 & 0 & 0 \end{bmatrix}
+\begin{bmatrix} 0 & 0 & 0 & 0 & 0 & 0 & 0 & -1 \\ 0 & 0 & 0 & 0 & 0 & 0 & 0 & 1 \\ 0 & 0 & 0 & 0 & 0 & -1 & 0 & 0 \\ 0 & 0 & 0 & 0 & 0 & 1 & 0 & 0 \\ 0 & 0 & 0 & 1 & 0 & 0 & 0 & 0 \\ 0 & 0 & 0 & -1 & 0 & 0 & 0 & 0 \\ 0 & 1 & 0 & 0 & 0 & 0 & 0 & 0 \\ 0 & -1 & 0 & 0 & 0 & 0 & 0 & 0 \end{bmatrix}
$$

|        | $H0_{8R}$ | $H1_{8R}$ | $H2_{8R}$ | $H3_{8R}$ |
|--------|-----------|-----------|-----------|-----------|
| $H0_{8R}$ | $H0_{8R}$ | $H1_{8R}$ | $H2_{8R}$ | $H3_{8R}$ |
| $H1_{8R}$ | $H1_{8R}$ | $-H0_{8R}$ | $H3_{8R}$ | $-H2_{8R}$ |
| $H2_{8R}$ | $H2_{8R}$ | $-H3_{8R}$ | $-H0_{8R}$ | $H1_{8R}$ |
| $H3_{8R}$ | $H3_{8R}$ | $H2_{8R}$ | $-H1_{8R}$ | $-H0_{8R}$ |

Figure 48. Upper rows: the decomposition of the matrix $H_{8R}$ (from Figure 46) as sum of 4 matrices: $H_{8R} = H0_{8R} + H1_{8R} + H2_{8R} + H3_{8R}$. Bottom row: the multiplication table of these 4 matrices $H0_{8R}$, $H1_{8R}$, $H2_{8R}$ and $H3_{8R}$, which is identical to the multiplication table of quaternions by Hamilton. $H0_{8R}$ is identity matrix for this matrix set and it plays the role of the real unit here.

The initial (8*8)-matrix $H_8$ (Figure 41) can be also decomposed in another way, or more precisely in accordance with the structure of the (8*8)-matrix of dyadic shifts from Figure 1. Figure 49 shows the case of such dyadic-shift decomposition $H_8 = H0_8+H1_8+H2_8+H3_8+H4_8+H5_8+H6_8+H7_8$, when 8 sparse matrices $H0_8$, $H1_8$, $H2_8$, $H3_8$, $H4_8$, $H5_8$, $H6_8$, $H7_8$ arise ($H0_8$ is identity matrix for this matrix set). The set $H0_8$, $H1_8$, $H2_8$, $H3_8$, $H4_8$, $H5_8$, $H6_8$, $H7_8$ is closed in relation to multiplication and it defines the multiplication table on Figure 39. This multiplication table is identical to the multiplication table of 8-dimensional hypercomplex numbers, which are termed as biquaternions by J.Hamilton (or Hamiltons' quaternions over the field of complex numbers). General expression for biquaternions in this case can be written as $Q_8 = a_0*H0_8+a_1*H1_8+a_2*H2_8+a_3*H3_8+a_4*H4_8+a_5*H5_8+a_6*H6_8+a_7*H7_8$, where $a_0, a_1, a_2, a_3, a_4, a_5, a_6, a_7$ are real numbers. From this point of view, the (8*8)-genomatrix $H_8$ is Hamiltons' bi-quaternion with unit coordinates.

$H_8 = H0_8+H1_8+H2_8+H3_8+H4_8+H5_8+H6_8+H7_8 =$

$$
\begin{bmatrix} 1 & 0 & 0 & 0 & 0 & 0 & 0 & 0 \\ 0 & 1 & 0 & 0 & 0 & 0 & 0 & 0 \\ 0 & 0 & 1 & 0 & 0 & 0 & 0 & 0 \\ 0 & 0 & 0 & 1 & 0 & 0 & 0 & 0 \\ 0 & 0 & 0 & 0 & 1 & 0 & 0 & 0 \\ 0 & 0 & 0 & 0 & 0 & 1 & 0 & 0 \\ 0 & 0 & 0 & 0 & 0 & 0 & 1 & 0 \\ 0 & 0 & 0 & 0 & 0 & 0 & 0 & 1 \end{bmatrix}
+\begin{bmatrix} 0 & -1 & 0 & 0 & 0 & 0 & 0 & 0 \\ 1 & 0 & 0 & 0 & 0 & 0 & 0 & 0 \\ 0 & 0 & 0 & -1 & 0 & 0 & 0 & 0 \\ 0 & 0 & 1 & 0 & 0 & 0 & 0 & 0 \\ 0 & 0 & 0 & 0 & 0 & -1 & 0 & 0 \\ 0 & 0 & 0 & 0 & 1 & 0 & 0 & 0 \\ 0 & 0 & 0 & 0 & 0 & 0 & 0 & -1 \\ 0 & 0 & 0 & 0 & 0 & 0 & 1 & 0 \end{bmatrix}
+\begin{bmatrix} 0 & 0 & 1 & 0 & 0 & 0 & 0 & 0 \\ 0 & 0 & 0 & 1 & 0 & 0 & 0 & 0 \\ -1 & 0 & 0 & 0 & 0 & 0 & 0 & 0 \\ 0 & -1 & 0 & 0 & 0 & 0 & 0 & 0 \\ 0 & 0 & 0 & 0 & 0 & 1 & 0 & 0 \\ 0 & 0 & 0 & 0 & 0 & 0 & 0 & 1 \\ 0 & 0 & 0 & 0 & -1 & 0 & 0 & 0 \\ 0 & 0 & 0 & 0 & 0 & -1 & 0 & 0 \end{bmatrix}
+\begin{bmatrix} 0 & 0 & 0 & -1 & 0 & 0 & 0 & 0 \\ 0 & 0 & 1 & 0 & 0 & 0 & 0 & 0 \\ 0 & 1 & 0 & 0 & 0 & 0 & 0 & 0 \\ -1 & 0 & 0 & 0 & 0 & 0 & 0 & 0 \\ 0 & 0 & 0 & 0 & 0 & 0 & 0 & -1 \\ 0 & 0 & 0 & 0 & 0 & 0 & 1 & 0 \\ 0 & 0 & 0 & 0 & 0 & 1 & 0 & 0 \\ 0 & 0 & 0 & 0 & -1 & 0 & 0 & 0 \end{bmatrix}+
$$

$$
\begin{bmatrix} 0 & 0 & 0 & 0 & -1 & 0 & 0 & 0 \\ 0 & 0 & 0 & 0 & 0 & -1 & 0 & 0 \\ 0 & 0 & 0 & 0 & 0 & 0 & 1 & 0 \\ 0 & 0 & 0 & 0 & 0 & 0 & 0 & 1 \\ 1 & 0 & 0 & 0 & 0 & 0 & 0 & 0 \\ 0 & 1 & 0 & 0 & 0 & 0 & 0 & 0 \\ 0 & 0 & -1 & 0 & 0 & 0 & 0 & 0 \\ 0 & 0 & 0 & -1 & 0 & 0 & 0 & 0 \end{bmatrix}
+\begin{bmatrix} 0 & 0 & 0 & 0 & 0 & 1 & 0 & 0 \\ 0 & 0 & 0 & 0 & -1 & 0 & 0 & 0 \\ 0 & 0 & 0 & 0 & 0 & 0 & 0 & -1 \\ 0 & 0 & 0 & 0 & 0 & 0 & 1 & 0 \\ 0 & -1 & 0 & 0 & 0 & 0 & 0 & 0 \\ 1 & 0 & 0 & 0 & 0 & 0 & 0 & 0 \\ 0 & 0 & 0 & 1 & 0 & 0 & 0 & 0 \\ 0 & 0 & -1 & 0 & 0 & 0 & 0 & 0 \end{bmatrix}
+\begin{bmatrix} 0 & 0 & 0 & 0 & 0 & 0 & 1 & 0 \\ 0 & 0 & 0 & 0 & 0 & 0 & 0 & 1 \\ 0 & 0 & 0 & 0 & 1 & 0 & 0 & 0 \\ 0 & 0 & 0 & 0 & 0 & 1 & 0 & 0 \\ 0 & 0 & -1 & 0 & 0 & 0 & 0 & 0 \\ 0 & 0 & 0 & -1 & 0 & 0 & 0 & 0 \\ -1 & 0 & 0 & 0 & 0 & 0 & 0 & 0 \\ 0 & -1 & 0 & 0 & 0 & 0 & 0 & 0 \end{bmatrix}
+\begin{bmatrix} 0 & 0 & 0 & 0 & 0 & 0 & 0 & -1 \\ 0 & 0 & 0 & 0 & 0 & 0 & 1 & 0 \\ 0 & 0 & 0 & 0 & 0 & -1 & 0 & 0 \\ 0 & 0 & 0 & 0 & 1 & 0 & 0 & 0 \\ 0 & 0 & 0 & 1 & 0 & 0 & 0 & 0 \\ 0 & 0 & -1 & 0 & 0 & 0 & 0 & 0 \\ 0 & 1 & 0 & 0 & 0 & 0 & 0 & 0 \\ -1 & 0 & 0 & 0 & 0 & 0 & 0 & 0 \end{bmatrix}
$$

|        | 1      | $H1_8$  | $H2_8$  | $H3_8$  | $H4_8$  | $H5_8$  | $H6_8$  | $H7_8$  |
|--------|--------|---------|---------|---------|---------|---------|---------|---------|
| **1**  | 1      | $H1_8$  | $H2_8$  | $H3_8$  | $H4_8$  | $H5_8$  | $H6_8$  | $H7_8$  |
| $H1_8$ | $H1_8$ | **-1**  | $H3_8$  | $-H2_8$ | $H5_8$  | $-H4_8$ | $H7_8$  | $-H6_8$ |
| $H2_8$ | $H2_8$ | $H3_8$  | **-1**  | $-H1_8$ | $-H6_8$ | $-H7_8$ | $H4_8$  | $H5_8$  |
| $H3_8$ | $H3_8$ | $-H2_8$ | $-H1_8$ | **1**   | $-H7_8$ | $H6_8$  | $H5_8$  | $-H4_8$ |
| $H4_8$ | $H4_8$ | $H5_8$  | $H6_8$  | $H7_8$  | **-1**  | $-H1_8$ | $-H2_8$ | $-H3_8$ |
| $H5_8$ | $H5_8$ | $-H4_8$ | $H7_8$  | $-H6_8$ | $-H1_8$ | **1**   | $-H3_8$ | $H2_8$  |
| $H6_8$ | $H6_8$ | $H7_8$  | $-H4_8$ | $-H5_8$ | $H2_8$  | $H3_8$  | **-1**  | $-H1_8$ |
| $H7_8$ | $H7_8$ | $-H6_8$ | $-H5_8$ | $H4_8$  | $H3_8$  | $-H2_8$ | $-H1_8$ | **1**   |

Figure 49. Upper rows: the decomposition of the matrix $H_8$ (from Figure 41) as sum of 8 matrices: $H_8 = H0_8 + H1_8 + H2_8 + H3_8 + H4_8 + H5_8 + H6_8 + H7_8$. Bottom row: the multiplication table of these 8 matrices $H0_8$, $H1_8$, $H2_8$, $H3_8$, $H4_8$, $H5_8$, $H6_8$ and $H7_8$, which is identical to the multiplication table of biquaternions by Hamilton (or Hamiltons' quaternions over the field of complex numbers). $H0_8$ is identity matrix.

Here for the (8*8)-genomatrix H8 we have received the interesting result: the sum of two different 4-dimensional quaternions by Hamilton with unit coordinates (they belong to two different matrix representations of Hamiltons' quaternions) generates the 8-dimensional biquaternion with unit coordinates. This result resembles results, which were described above in this section about genetic matrices with internal complementarities, and it resembles again a situation when a union of Yin and Yang (a union of male and female beginnings, or a fusion of male and female gametes) generates a new organism.

Phenomenology of the genetic system gives additional confirmations of its connection with the mosaic genomatrices [C T; A G]$^{(n)}$, numeric representations of which posses internal complementarities (Figure 41). In matrices [C T; A G]$^{(n)}$, let us numerate their $2^n$ columns from left to right by numbers 0, 1, 2, .., $2^n$-1 and then consider two sets of n-plets (oligonucleotides) in each of matrices [C T; A G]$^{(n)}$ (for example, for n=2, 3):
- The first set contains all n-plets from columns with even numeration 0, 2, 4, … (this set is conditionally termed as the even-set or the Yin-set);
- The second set contains all n-plets from columns with odd numeration 1, 3, 5, … (this set is conditionally termed as the odd-set or the Yang-set).

For example, the genomatrix [C T; A G]$^{(3)}$ (Figure 41) contains the even-set of 32 triplets in its columns with even numerations 0, 2, 4, 6 (CCC, CCA, CAC, CAA, ACC, ACA, AAC, AAA, CTC, CTA, CGC, CGA, ATC, ATA, AGC, AGA, TCC, TCA, TAC, TAA, GCC, GCA, GAC, GAA, TTC, TTA, TGC, TGA, GTC, GTA, GGC, GGA) and the odd-set of 32 triplets in its columns with odd numerations 1, 3, 5, 7 (CCT, CCG, CAT, CAG, ACT, ACG, AAT, AAG, CTT, CTG, CGT, CGG, ATT, ATG, AGT, AGG, TCT, TCG, TAT, TAG, GCT, GCG, GAT, GAG, TTT, TTG, TGT, TGG, GTT, GTG, GGT, GGG). One can show, for example, that the structure of the whole human genome is connected with the devision of the whole set of 64 triplets into the even-set of 32 triplets and the odd-set of 32 triplets. Really, let us calculate total quantities (frequencies $F_{even}$ and $F_{odd}$) of members of these two sets of triplets in the whole human genome, which contains the huge number 2.843.411.612 (about three billion) triplets and data about which are shown above on Figure 5. The result of this calculation shows that the total quantity of members of the even-set ($F_{even}$) and the total quantity of members of the odd-set ($F_{odd}$) are equal to within 0,12%:
- $F_{even}$ = **1.420.853.821** for the even-set of 32 triplets (CCC, CCA, CAC, CAA, ACC, ACA, AAC, AAA, CTC, CTA, CGC, CGA, ATC, ATA, AGC, AGA, TCC, TCA, TAC, TAA, GCC, GCA, GAC, GAA, TTC, TTA, TGC, TGA, GTC, GTA, GGC, GGA);

- $F_{odd} = \mathbf{1.422.557.791}$ for the odd-set of 32 triplets (CCT, CCG, CAT, CAG, ACT, ACG, AAT, AAG, CTT, CTG, CGT, CGG, ATT, ATG, AGT, AGG, TCT, TCG, TAT, TAG, GCT, GCG, GAT, GAG, TTT, TTG, TGT, TGG, GTT, GTG, GGT, GGG);
- The percentage difference between these $F_{even}$ and $F_{odd}$ is equal to 0,12%.

More general confirmation of genetic importance of the structure of genomatrices with internal complementarities (where all set of n-plets is divided in sub-sets of n-plets from columns with even and odd numerations) is given by results of the study of the Symmetry Principle № 6 from the work [Petoukhov, 2008c, 6$^{th}$ version, section 11], where a notion of fractal genetic nets for long nucleotide sequences is used.

Described genetic matrices with internal complementarities posess many interesting mathematical properties, concerning with cyclic and dyadic shifts, multiplications of these matrices, Kronecker families of matrices $R_4 \otimes [1\ 1;\ 1\ 1]^{(n)}$ and $H_4 \otimes [1\ -1;\ 1\ 1]^{(n)}$, dyadic trees of different $2^n$-dimensional numbers, rotational transformations of these numeric genomatrices into new numeric genomatrices with internal complementarities, etc. A set of $(2^n*2^n)$-matrices with internal complementarities contains a huge quantity of different types of matrix representations of complex numbers (that is algebraic fields (http://en.wikipedia.org/wiki/Field_(mathematics))) and of split-complex numbers, which were previously unknown in mathematics, as the author can judge. The systems of $2^n$-dimensional numeric systems, which are described in this section, have perspectives to be applied in mathematical natural sciences and signals processing. The discovery of genetic importance of matrices with internal complementarities gives a possibility to divide sets of amino acids and stop-signals in interesting sub-sets in accordance with the structure of the genomatrix $[C\ T;\ A\ G]^{(3)}$; it gives also new approaches to study proteins. The author will describe these properties and materials in additional publications. Results of matrix genetics lead to the idea that the structure of the genetic coding system is dictated by patterns of described numeric genomatrices; here one can remind the famous Pythagorean statement that "numbers rule the world" with the refinement that we should talk now about multi-dimensional numbers.

**11. Conclusion**

This article puts forward the conception of "multi-lingual genetics" (or "multi-alphabetical genetics") for a discussion (see the section 1 about this). To this conception we have also presented materials about relations of some kinds of hypercomplex numbers, Fibonacci matrices and screw theory with the algebraic multi-level system of genetic alphabets and with inherited biological phenomena including phyllotaxis (in addition to our publications [Petoukhov, 2008a,b, 2011a,b, 2012; Petoukhov, He, 2010]). Some analogies between linguistics and molecular genetics were also considered.

The multi-level system of genetic alphabets is endowed with algebraic properties, which allows studying and modeling the genetic system in the language of the basic concept of mathematics - the concept of number. The use of hypercomplex numbers and their multi-dimensional vector spaces for the study of genetic system and heritable biological phenomena allows developing an algebraic biology for including biology in the area of advanced mathematical natural science. Genetic meanings of these hypercomplex numbers and algebraic properties of genetic alphabets are attentively been studying now in the author's laboratory.

Returning to dyadic-shift matrices (Figure 1) one can note their cross-wise character connected with their two diagonals: dispositions of members of dyadic groups inside each of matrices are identical in both quadrants along each diagonal. But genetical inherited constructions of physiological systems (including sensory-motion systems) demonstrate similar cross-wise structures by unknown reasons. For example, the connection between the hemispheres of human brain and the halves of human body possesses the similar cross-wise character: the left hemisphere serves the right half of the body and the right hemisphere (Figure 50) [Annett, 1985, 1992; Gazzaniga, 1995; Hellige, 1993]. The system of optic cranial nerves from two eyes possesses the cross-wise structures as well: the optic nerves transfer information about the right half of field of vision into the left hemisphere of brain, and information about the left half of field of vision into the right hemisphere. The same is held true for the hearing system [Penrose, 1989, Chapter 9].

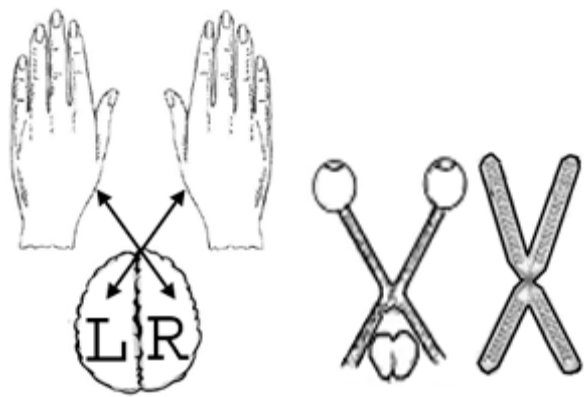

Figure 50. The cross-wise schemes of some morpho-functional structures in human organism. On the left side: the cross-wise connections of brain hemispheres with the left and the right halves of a human body. In the middle: the cross-wise structure of optic nerves from eyes in brain. On the right side: a chromosome (from [Petoukhov, He, 2010])

One can suppose that these genetically inherited phenomena of crosswise structures are connected with cross-wise matrices of dyadic shifts. Matrix genetics gives evidences in favor of the thesis: "We are the dyadic-shift organisms".

Results of matrix genetics lead to the idea that the structure of the genetic coding system is dictated by patterns of described numeric genomatrices; here one can remind the famous Pythagorean statement that "numbers rule the world" with the refinement that we should talk now about multi-dimensional numbers.

**Acknowledgments**. Described researches were made by the author in the frame of a long-term cooperation between Russian and Hungarian Academies of Sciences and in the frames of programs of "International Symmetry Association" (Hungary, http://symmetry.hu/) and of "International Society of Symmetry in Bioinformatics" (USA, http://polaris.nova.edu/MST/ISSB). The author is grateful to Darvas G., Ganiev R.F., He M., Adamson G., Cristea P., Kappraff J., Kulakov Y.I., Pavlov D.G., Pellionisz A.J., Vladimirov Y.S., Stepanyan I.V., Svirin V.I. for their support.